\documentclass[lettersize,journal]{IEEEtran}
\usepackage{amsmath,amsfonts}
\usepackage{algorithmic}
\usepackage{algorithm}
\usepackage{array}
\usepackage{subcaption}
\usepackage{textcomp}
\usepackage{stfloats}
\usepackage{url}
\usepackage{verbatim}
\usepackage{graphicx}
\usepackage{cite}
\usepackage{xcolor}
\usepackage{multirow}
\begin{document}

\title{Neural Network Training on In-memory-computing Hardware with Radix-4 Gradients}

\author{Christopher Grimm,~\IEEEmembership{Student Member,~IEEE,} and Naveen Verma,~\IEEEmembership{Member,~IEEE,}

\thanks{Christopher Grimm and Naveen Verma are with the Department of Electrical and Computer Engineering, Princeton University, Princeton, NJ 08544 USA (email: cgrimm@princeton.edu).}

\thanks{This work has been submitted to the IEEE for possible publication. Copyright may be transferred without notice, after which this version may no longer be accessible.}
}

\markboth{Journal of \LaTeX\ Class Files,~Vol.~14, No.~8, August~2021}%
{Shell \MakeLowercase{\textit{et al.}}: A Sample Article Using IEEEtran.cls for IEEE Journals}

\IEEEpubid{DOI 10.1109/TCSI.2022.3185556~\copyright~2022 IEEE}

\maketitle

\begin{abstract}
Deep learning training involves a large number of operations, which are dominated by high dimensionality Matrix-Vector Multiplies (MVMs). This has motivated hardware accelerators to enhance compute efficiency, but where data movement and accessing are proving to be key bottlenecks. In-Memory Computing (IMC) is an approach with the potential to overcome this, whereby computations are performed in-place within dense 2-D memory. However, IMC fundamentally trades efficiency and throughput gains for dynamic-range limitations, raising distinct challenges for training, where compute precision requirements are seen to be substantially higher than for inferencing. This paper explores training on IMC hardware by leveraging two recent developments: (1) a training algorithm enabling aggressive quantization through a radix-4 number representation; (2) IMC leveraging compute based on precision capacitors, whereby analog noise effects can be made well below quantization effects.
Energy modeling calibrated to a measured silicon prototype implemented in 16nm CMOS shows that energy savings of over 400$\times$ can be achieved with full quantizer adaptability, where all training MVMs can be mapped to IMC, and 3$\times$ can be achieved for two-level quantizer adaptability, where two of the three training MVMs can be mapped to IMC.
\end{abstract}


\section{Introduction}
\label{intro}

The success of deep learning in a broad range of tasks, from vision, to speech, to gameplay, and beyond \cite{alexnet, resnet, rnn, bert, alphago, diabdeep} has motivated its deployment in a vast array of applications and associated system platforms. However, deep learning involves a large and increasing number of computations in both the training and inferencing phases. In order to make its wide deployment feasible, much attention has turned to hardware acceleration for executing computations with greater energy efficiency and speed \cite{tn, tpu, rapid,tcas1,tcas2}.   

A critical consideration is that deep learning computations involve large amounts of data, making data movement and accessing a key bottleneck. Specifically, the dominate computation involved is high-dimensionality Matrix-Vector Mutliplies (MVMs). This has motivated two approaches to hardware acceleration. First is the use of spatial architectures (such as systolic arrays). Spatial architectures arrange compute and storage units in a 2-D array to exploit the data reuse inherent in MVMs. Second is the use of aggressive operand quantization which reduces both the computation and data-movement costs. Quantization has been widely explored for both training and inferencing, with studies showing that training is substantially more sensitive to bit precision and number representation format \cite{lim_prec, banner,rapid}.

\IEEEpubidadjcol
Regarding spatial architectures, an approach that has recently received much attention is In-Memory Computing (IMC). IMC takes the extreme approach of reducing compute and storage units to the dense, highly-parallel 2-D arrangement of bit cells within a memory. This typically requires exploiting analog operation to fit computation within the constrained bit cells. By doing so, IMC has shown the potential for enhancing both energy efficiency and throughput by over an order of magnitude \cite{tcas3,700TOPS}. However, IMC fundamentally institutes a trade-off where energy/throughput gains are derived at the cost of dynamic range, which is ultimately limited by analog noise effects. This has restricted the success of IMC demonstrations to inferencing, leaving training, which poses urgent challenges because it generally involves many more operations, largely unaddressed.      

In this work we aim to address the challenges with IMC training. We do this by exploiting two recent advances. First is a low-noise approach to IMC based on precision capacitors, which has enabled increased output bit precisions and suppression of analog noise effects well below the corresponding quantization level \cite{hossein_paper, 65nm}. Second is an aggressive approach to quantized training based on a radix-4 number format for the gradients \cite{rad4}. The key contributions of this work are as follows:     

\begin{enumerate}
\item We propose a method of mapping the radix-4 gradient format to XNOR computations within IMC hardware using a 1-hot encoded exponent. For training MVMs involving the gradients, this mapping leads to high levels of sparsity which directly reduces the dynamic-range requirements.
\item We propose a method for IMC hardware to practically exploit the dynamic-range reduction and identify the ultimate limitations that would be faced.
\item We develop TensorFlow training libraries corresponding to the specific quantization effects of the IMC hardware. This enables IMC-based training to be analyzed at different points in order to observe impacts of the proposed methods. 
\item We develop an approach for analyzing the benefits of IMC training hardware, based on energy models derived from silicon measurements of an IMC test chip in 16nm CMOS. We use this to evaluate the potential gains of IMC-based training.
\end{enumerate}

\section{Related Work}

\subsection{Quantized Training Methods}

Quantized training methods can be split into two broad categories: quantization-aware training and quantized training. In quantization-aware training only the forward MVM associated with inferencing is quantized while the training-specific MVMs are kept at a higher precision such as 32-bit floating point (FP32). This creates a more efficient model during inferencing. Binarized neural networks and wide-reduced precision networks represent two prominent examples associated with quantization-aware training, through which similar testing accuracies as FP32 models has been achieved while reducing activations and weights to as low as 1-bit resolution \cite{bin, wrpn}.

In quantized training, one or both of the training-specific MVMs are also quantized, enabling efficiency benefits during training. Early work showed that using stochastic rounding to prevent rounding bias could enable 16-bit fixed-precision values for all three training MVMs \cite{lim_prec}. Further work in integer-based training was able to show that 8-bit integer activations, weights, and gradients could be used for the forward and backward MVMs while the weight-update MVM was kept at FP32 \cite{banner}. \cite{wage} demonstrates a quantized training method which uses a tertiary quantization for the weights, and an 8-bit fixed point integer representation for the activations, gradients, and errors. To maximize the energy efficiency of the training approach, this method minimizes the number of floating-point operations by replacing batch normalization with a simple scaling approach and uses stochastic gradient descent without momentum.

Recently, new methods using reduced-bit floating-point and radix formats for the gradients have achieved significant success thanks to extended dynamic range. A method using 8-bit floating-point operands for the forward, backward, and weight-update MVMs demonstrated no loss in precision on a variety of Deep Neural Networks (DNNs) \cite{8bit}. An even more aggressive approach, which is central to the work in this paper, extends DNN training to 4-bit values for all three training MVMs. In \cite{rad4} radix-4 4-bit floating-point gradients are used along with 4-bit integer activations and weights. This gradient format is composed of a single sign bit and 3 exponent bits for representing the power of a base-4 exponent. This format permits a symmetric representation of gradients as large as $64$ and as small as $\frac{1}{64}$. Furthermore, radix-4 gradients are augmented by gradscaling as outlined in \cite{rad4}, which allows the appropriate scaling factor for the gradients to be learned during training to ensure maximal use of the extended dynamic range.

\subsection{In-Memory Computing}

IMC addresses throughput and energy of data-movement through a fundamental trade-off with dynamic range \cite{imc_reg}. Consider an $R$-row and $C$-column memory bank assumed to store the data required for a computation. Conventional memory accessing would require $R$ cycles, addressing each row one at a time using world-line (WL) signals, and accessing each row's bit cells in parallel via perpendicular bit-line (BL) signals. Instead, IMC accesses \emph{multiple} rows at once: providing input data on the WLs; performing computation (typically multiplication) with stored data in the bit cells; and then performing reduction (typically accumulation) on the column BLs. What is thus communicated is a compute result over many inputs and bit cells, thereby reducing the communication energy and delay by an amount directly related to the row parallelism. However, the compute result also requires higher dynamic range (i.e., compared to communicating individual bits), which generally also increases with the row parallelism (i.e., dimensionality of accumulation). Thus, energy and throughput gains come at the cost of dynamic range.       

In practice, the dynamic range of IMC is typically limited by noise sources (variations, non-idealities) due to analog operation in the bit cells. Recently, a capacitor-based approach to analog computation in IMC has demonstrated substantial reduction of noise \cite{hossein_paper}. This capacitor-based approach enables high output bit precisions (beyond 8-bit) and exhibits analog noise effects well below the output quantization level \cite{65nm}. These benefits have enabled the creation of robust abstractions in SRAM memories for integration and scale-up in digital architectures \cite{16nm}. Our work builds on these advances, together with the advances in quantized training, to explore the potential gains of IMC for training systems.

\section{Mapping Radix-4 based training operations to IMC}
\label{sec:rad4}

\begin{figure}[htb]
\centering
\includegraphics[width=1\linewidth]{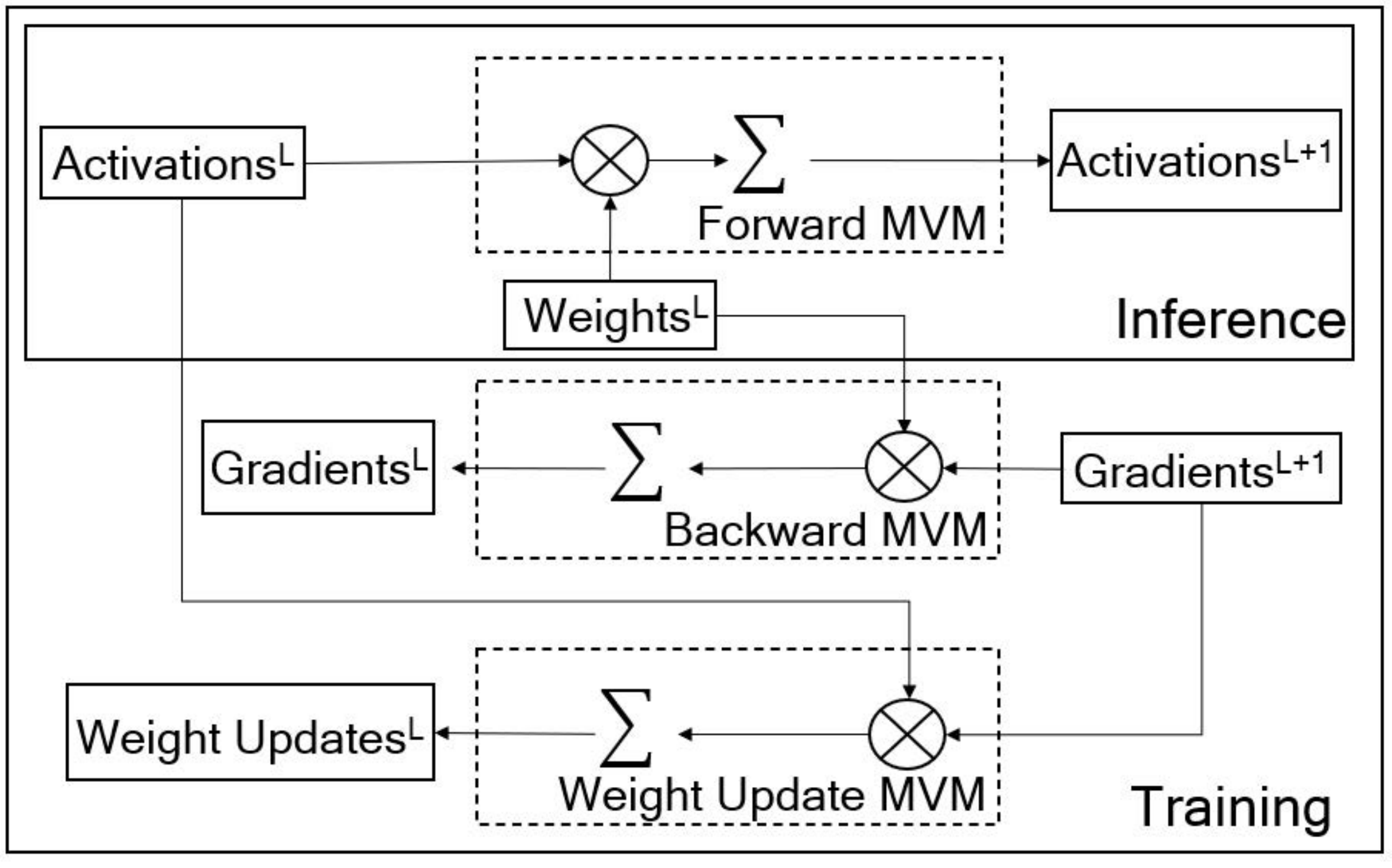}
\caption{Training involves three kinds of MVMs (inferencing involves just one). Each of the training MVMs requires two of three matrix operands: activations, weights, and gradients.}
\label{fig:mvm_break}
\end{figure}

Figure \ref{fig:mvm_break} shows the MVMs involved in training and inferencing. The work presented in this paper leverages a radix-4 +/-1-binary format for gradients that enables IMC in the backward and weight update MVMs, while a traditional +/-1-binary format is used for the activations and weights. This adjustment to the backward and weight update MVMs creates low enough quantization noise to permit training with IMC for all three training MVMs. Each of the different MVMs exhibit varying compute precision requirements, but acceleration of all MVMs is necessary to achieve overall energy efficiency and throughput advantages.  

\subsection{Training Operations}
\label{subsec:train_ops}
Each of the three types of MVMs involves multiplication between two of the following three variables: activations, weights, and gradients. The three MVMs are: the forward MVM, where activations are multiplied by weights; the backward MVM, where gradients are multiplied by weights; and the weight-update MVM, where activations are multiplied by gradients. Each of these MVMs has different quantization requirements for the inputs. In \cite{banner}, it was shown that operand quantization for the backward pass requires higher bit precision than the forward pass, while \cite{8bit} evaluated the accumulation quantization requirements for each of the MVMs. It is these higher quantization requirments in the backward and weight-update MVMs which have generally prevented low-precision hardware acceleration for training.

While the underlying operation performed by accelerators (e.g., spatial architectures, including IMC) is MVMs, these are typically composed into matrix-matrix multiplications. Taking a dense layer as an example, an activation matrix has size of $ B \times M$, a weight matrix has size of $ M \times N$, and a gradient matrix has size of $B \times N$, where $B$ is the batch size, $M$ is the number of elements per input, and $N$ is the number of neurons or filters. Based on these matrix sizes, the forward MVM produces a $B \times N$ matrix, the backward MVM produces a $B \times M$ matrix, and the weight update MVM produces a $M \times N$ matrix. Though the sizes of the matrices vary at each layer, they depend on the sizes of the subsequent/preceding layer. Thus, the overall element multiply-accumulate (MAC) operations for each of the three MVMs is equal across the entire model. This underscores that while the three MVMs exhibit different susceptibility to quantization noise, acceleration of all three is critical for training efficiency and throughput. 

Beyond MVM operations, DNN training also includes scalar multiplications, additions, and various activation functions such as the Rectified Linear Activation Unit (ReLU). However, it has been shown that MVM operations typically compose well beyond 70\% of the operations \cite{profile_paper}. Furthermore, these operations apply to single elements, and thus benefit from traditional acceleration approaches without the challenges of data movement and accessing.

\subsection{Forward MVM mapping to IMC}
\label{subsec:mvm_for}

\begin{figure*}[htb]
\centering
\includegraphics[width=1\linewidth]{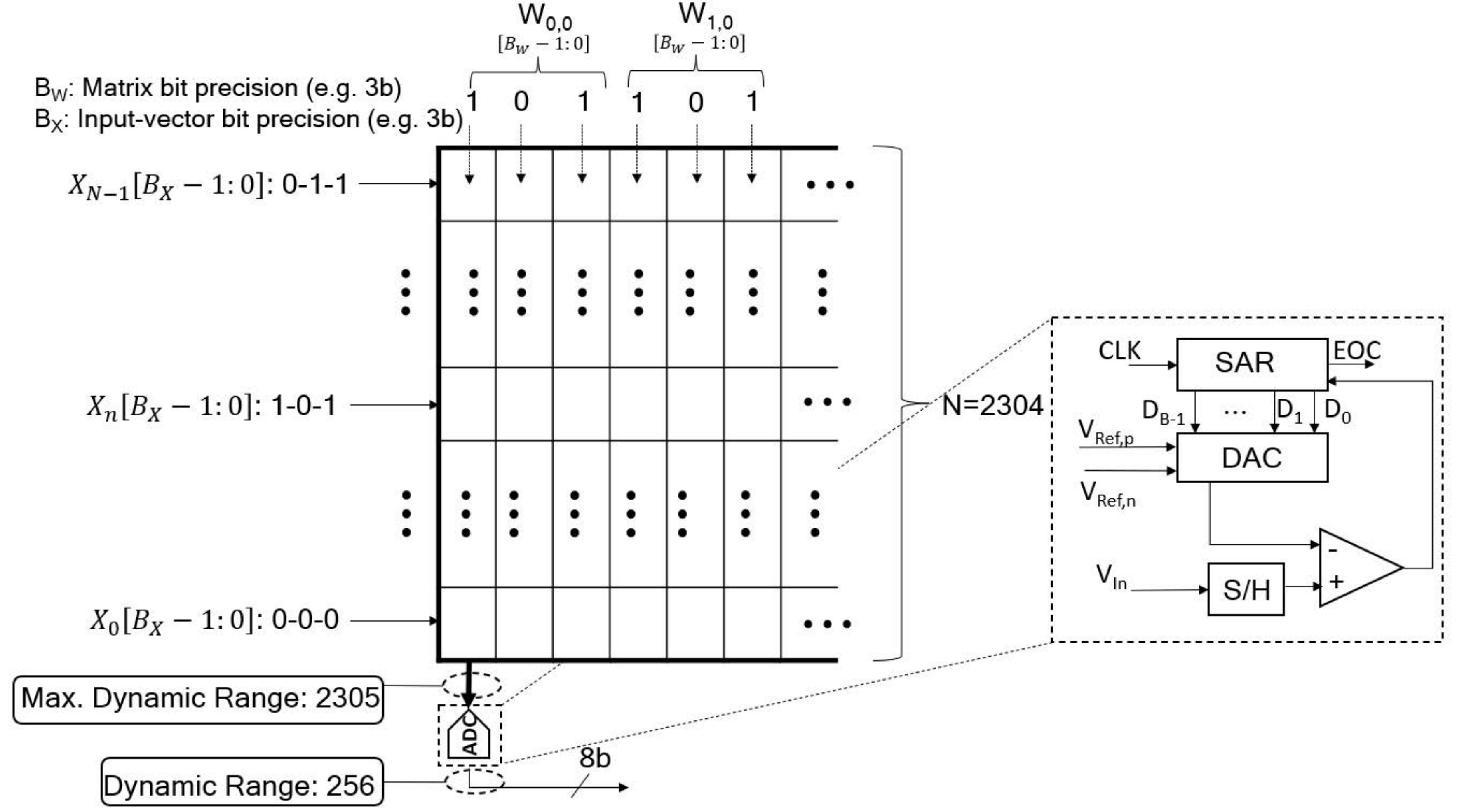}
\caption{MVM mapping to IMC hardware, based on BPBS approach, whereby column computations correspond to binary-vector inner products. The output ADC introduces quanitzation effects in BPBS reconstruction. A block diagram of the SAR ADC is included.}
\label{fig:xnor_form}
\end{figure*}

For the forward MVM, a traditional +/-1-binary format is mapped to IMC. Previous work has already demonstrated effective mapping of forward MVMs to IMC hardware. For this work, we assume the precise capacitor-based approach in \cite{65nm, 16nm}, with similar assumptions on the hardware: 2304 memory rows, 256 memory columns, and 8-bit output ADCs. The capacitor-based approach performs binary-vector inner products in each column. Extension to multi-bit elements is achieved through bit-parallel/bit-serial (BPBS) operation, as shown in Figure \ref{fig:xnor_form}, where multiple matrix-element bits are mapped to parallel columns, and multiple input-vector bits are provided serially. This yields binary operations in each column, and allows reconstruction of the multi-bit result through binary weighting (bit shifting) and summation across the digitized column outputs. As described in \cite{65nm} weights are quantized off-chip and loaded into the compute in memory array (CIMA) as 32-bit doublewords with each column composed of the values from a single filter of size $M$. Activations are quantized off-chip and then streamed into the input reshape buffer as 32-bit doublewords. The input reshape buffer properly aligns the activations and takes advantage of data reuse in convolutional layers. The activations are then fed to the CIMA where each bit-cell performs a multiplication operation and stored on a capacitor. All the capacitors along a column are shorted together and the resulting voltage is fed to an 8-bit Successive Approximation Register (SAR) ADC. The SAR logic initially sets the digital output $[D_{B-1},D_0]$ where only the Most Significant Bit (MSB) is set to one and this predicted output is fed into the Digital-to-Analog Converter (DAC) to generate the appropriate analog voltage between $V_{Ref,p}$ and $V_{Ref,n}$. This voltage is compared to the input voltage $V_{in}$ held in the Set-and-Hold (S/H) circuit. Based on the output of the comparator, the MSB is set to 0 if $V_{in}$ is less than the voltage generated by the DAC or kept at 1 otherwise. The next most significant bit is set to one and the process is repeated until the Least Significant Bit (LSB) has been set and the analog input voltage digitally converted. At this point the end of conversion (EOC) signal is set high, signifying the ADC digital output is ready. The resulting output is offset and scaled according to a variety of factors including the quantization and sparsity of the inputs by the near-memory computing (NMC) data path.

This IMC mapping reduces multiplication within the bit cells to binary XNOR operations, with inputs corresponding to +/-1. In \cite{65nm}, a +/-1 number-representation format is proposed, where a $K$-bit integer number $x$ is represented using the following equation:

\begin{equation}
\label{eqn:xnor_bin}
x = \sum_{i=1}^{K-2} b_i \times 2^{i-1} + (b_{0+} + b_{0-}) \times 2^{-1},
\end{equation}

where $b_i$ represents the +1 or -1 value at bit-position $i$. This representation generally has an integer value between $[\ -2^{K-2},  2^{K-2} ]\ $. 

While capacitor-based IMC has the advantage of suppressing analog noise effects well below the quantization level of the 8-bit output ADC, the ADC together with the BPBS approach introduces specific quantization noise. Namely, accumulation of binary XNOR outputs along a 2304-dimensional column results in levels corresponding to a range from $[0,2304]$, and thus with a dynamic range of 2305 levels. However, the 8-bit ADC, with resolution chosen to balance energy and area overheads, forces quantization to 256 levels. This dynamic-range mismatch introduces quantization noise within the BPBS reconstruction process, which must be analyzed for training-MVM accuracy.

\subsection{Radix-4 MVM mapping for IMC}
\label{subsec:rad4_mapping}

\begin{figure}[htb]
\centering
\includegraphics[width=1\linewidth]{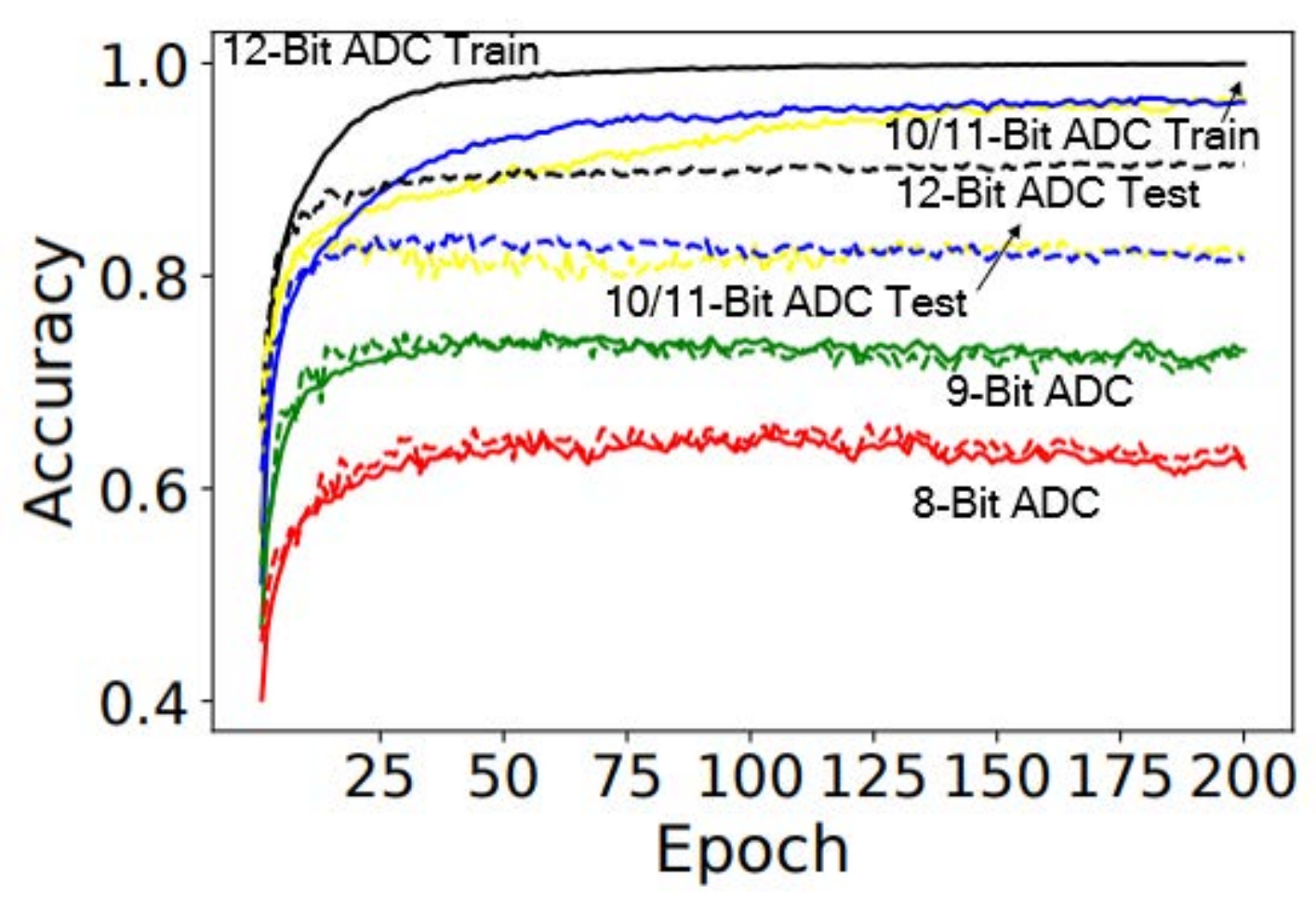}
\caption{Applying the 8-bit integer training approaches from \cite{banner} on a 9 layer VGG-lite model trained on the CIFAR-10 dataset using IMC quantization noise. This approach results in severe degradation of both training and test accuracy unless the ADC quantization noise is reduced by significantly increasing the ADC precision.}
\label{fig:imc_int}
\end{figure}

While it is possible to represent MVM input gradients using the format in Equation \ref{eqn:xnor_bin}, previous work has shown that at the 8-bit quantization level necessary for gradient precision, BPBS computation introduces notable additional quantization noise \cite{65nm}. We further demonstrate this in Figure \ref{fig:imc_int}, where using 8-bit integer activation, weights, and gradients for the forward and backward MVM operations as in \cite{banner} results in a significant degradation in accuracy when implemented under the ADC quantization noise constraints of IMC. Recovery of accuracy was achieved by increasing the ADC precision and greatly reducing the dynamic range mismatch between the CIMA column output and the ADC. Full recovery of both training and test accuracy was only possible by completely eliminating the dynamic range mismatch and using a 12-bit ADC which drastically reduces the efficiency of IMC. Thus, we focus on the radix-4 representation of gradients presented in \cite{rad4}. This work demonstrates a novel method for mapping radix-4 gradients to IMC by proposing a 1-hot encoding of the exponent bits. This yields a representation for $K$-bit gradients consisting of a single sign bit, and, at most, one non-zero exponent bit. This format can be represented by the following equation:

\begin{equation}
\label{eqn:rad_bin}
x = s \cdot \sum_{i=0}^{K-2} M_i \times 4^{i- \frac{K-2}{2}},
\end{equation}
where $s$ is the value of the sign bit, $M_i$ is the $i^{th}$ mask bit, and we assume $K$ is even.

As presented in \cite{rad4}, the radix-4 gradients appropriately capture the range of the log-normal distributed gradients. While a radix-2 method could capture the same range as the radix-4 method, radix-2 would require 13 instead of 7 1-hot encoded exponent bits which would greatly reduce the energy efficiency and latency of the method. Using a larger base such as radix-8 would capture a similar range in fewer bits, but fails to capture the necessary precision to perform training. It is for these reasons that the radix-4 format was chosen.

\begin{figure}[tb]
\centering
\includegraphics[width=1\linewidth]{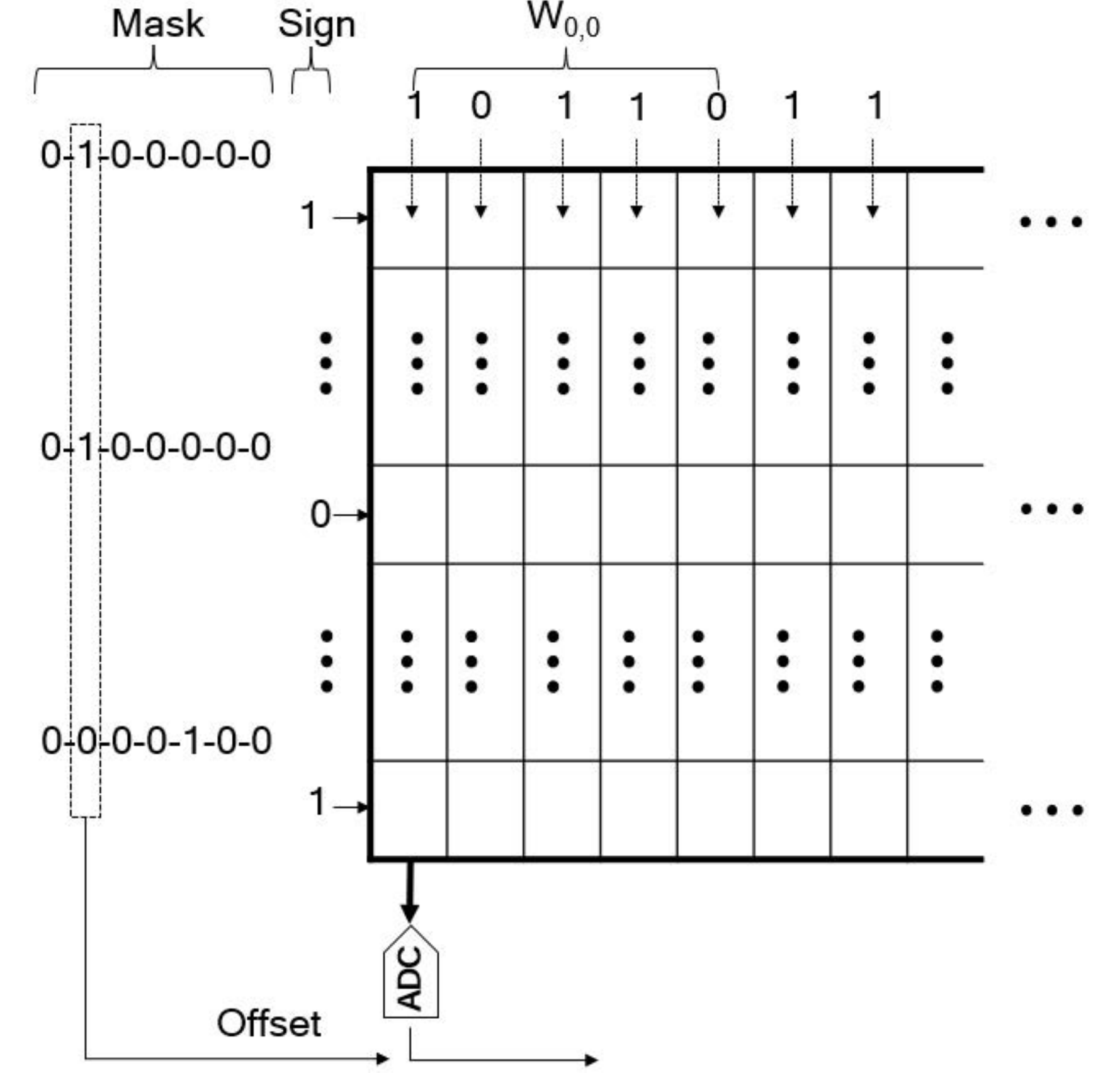}
\caption{Application of radix-4 inputs, with 1-hot encoded exponent representation, to IMC. The 1-hot encoded exponent bits serve as serially-applied mask bits for the input sign bit.}
\label{fig:radix_form}
\end{figure}

Mapping of the proposed radix-4 representation to IMC is illustrated in Figure \ref{fig:radix_form}. Only the sign bit is applied to the input while the exponent bits serve as a mask. In \cite{65nm}, it is described how a mask can be readily applied to prevent IMC XNOR computation for the row. The total number of masked bits can then be applied as a digital offset after the ADC to negate accumulation of the masked XNOR outputs. This ensures that each IMC operation only involves input elements with the same base-4 power. Subsequently, properly weighting and summing the serial IMC outputs across the base-4 input powers yields results that are mathematically equivalent to the radix-4-input MVM operation (except for the ADC quantization effects incurred).

For the backward MVMs, the weights are quantized off-chip and loaded into the CIMA as 32-bit doublewords with each column composed of the values from a single input of size $N$. Gradients are quantized and one-hot-encoded off-chip and then streamed into the input reshape buffer as 32-bit doublewords. The gradients are then fed to the CIMA where each bit-cell performs a multiplication operation and stores its result on a capacitor. All the capacitors along a column are shorted together and the resulting voltage is fed to an 8-bit ADC. The resulting digital output is offset and scaled according to a variety of factors including the quantization and sparsity of the gradients by the NMC data path. This process is repeated for the weight update MVM with the activations loaded into the CIMA with each column composed of values from a single input of size $B$ and the gradients again being streamed into the CIMA.

\begin{figure}[tb]
\centering
\includegraphics[width=1\linewidth]{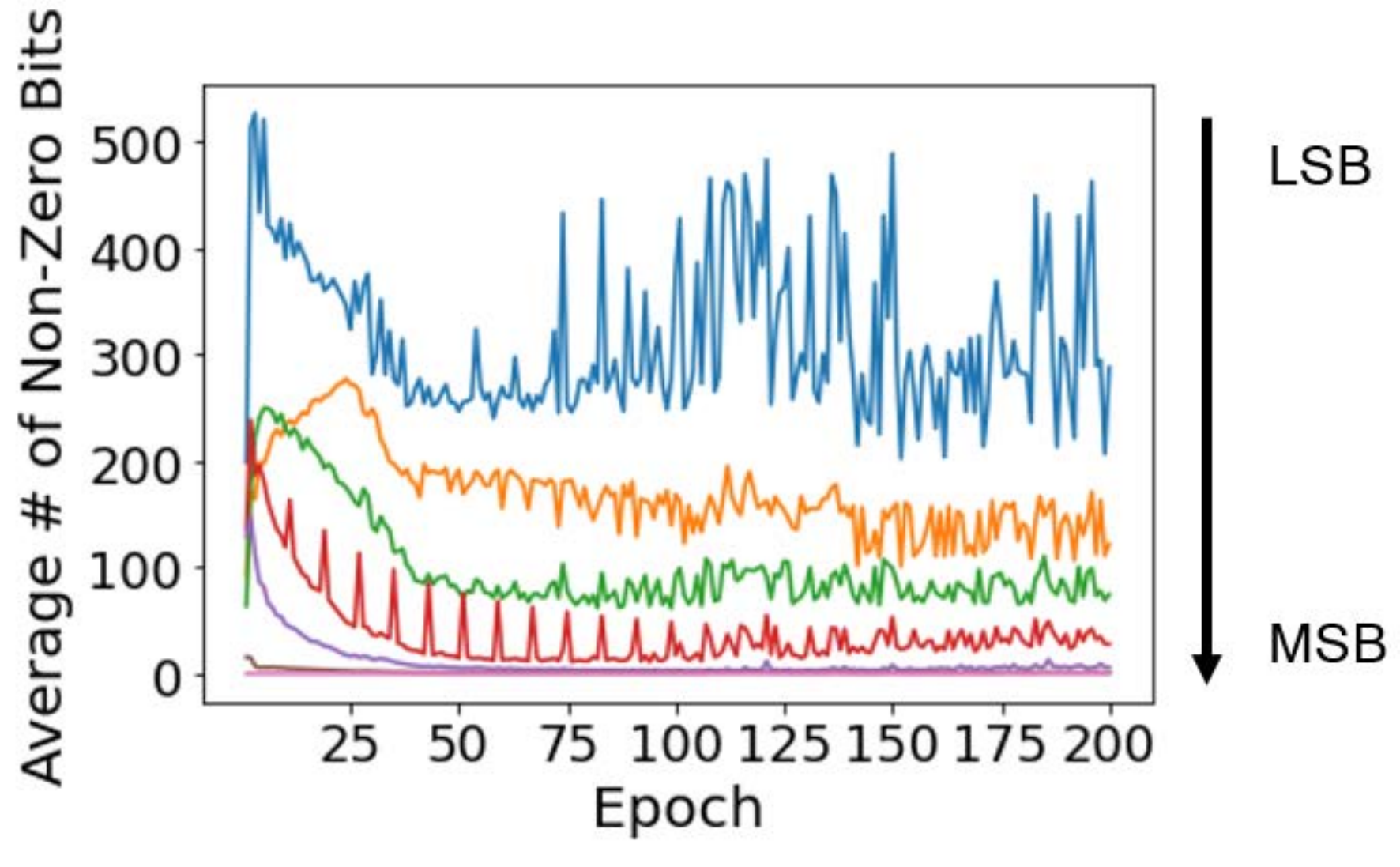}
\caption{The average number of active rows for each exponent bit in the fifth convolutional layer of a VGG-lite model. The $8^{th}$ bit is a sign bit and has no effect on the number of active rows.}
\label{fig:bits}
\end{figure}
While the 1-hot encoding of exponent bits increases the number of serial IMC operations, the low exponent precision makes this tolerable. Importantly, 1-hot encoding introduces two advantages. First, masking saves input-driving energy. Second, masking leads to high input sparsity, reducing the column dynamic range from 2304 to just the number of non-zero values. Since no more than a single value of the gradient mask is set to 1, this creates at most an average of $\frac{1}{K-1}$ active rows per input vector. In practice the number of active rows is actually less than this due to the large number of gradients with a value of 0, as shown in Figure \ref{fig:bits}. This figure plots the number of active rows for the fifth convolutional layer of a VGG-lite model (i.e., layer with the largest accumulation dynamic range and least sparse gradients). We can see that even the least sparse vector typically uses less than a fifth of the 2304 rows, and thus requires a fifth of the dynamic range. Furthermore, we see that the highest-order exponent bit (MSB) yields the greatest input sparsity, thus mitigating the dynamic range and quantization effects for the most critical IMC operation.

\subsection{Dynamic-range Adaptive IMC}
\label{subsec:dyn_range_imc}

\begin{figure}[tb]
\centering
\includegraphics[width=1\linewidth]{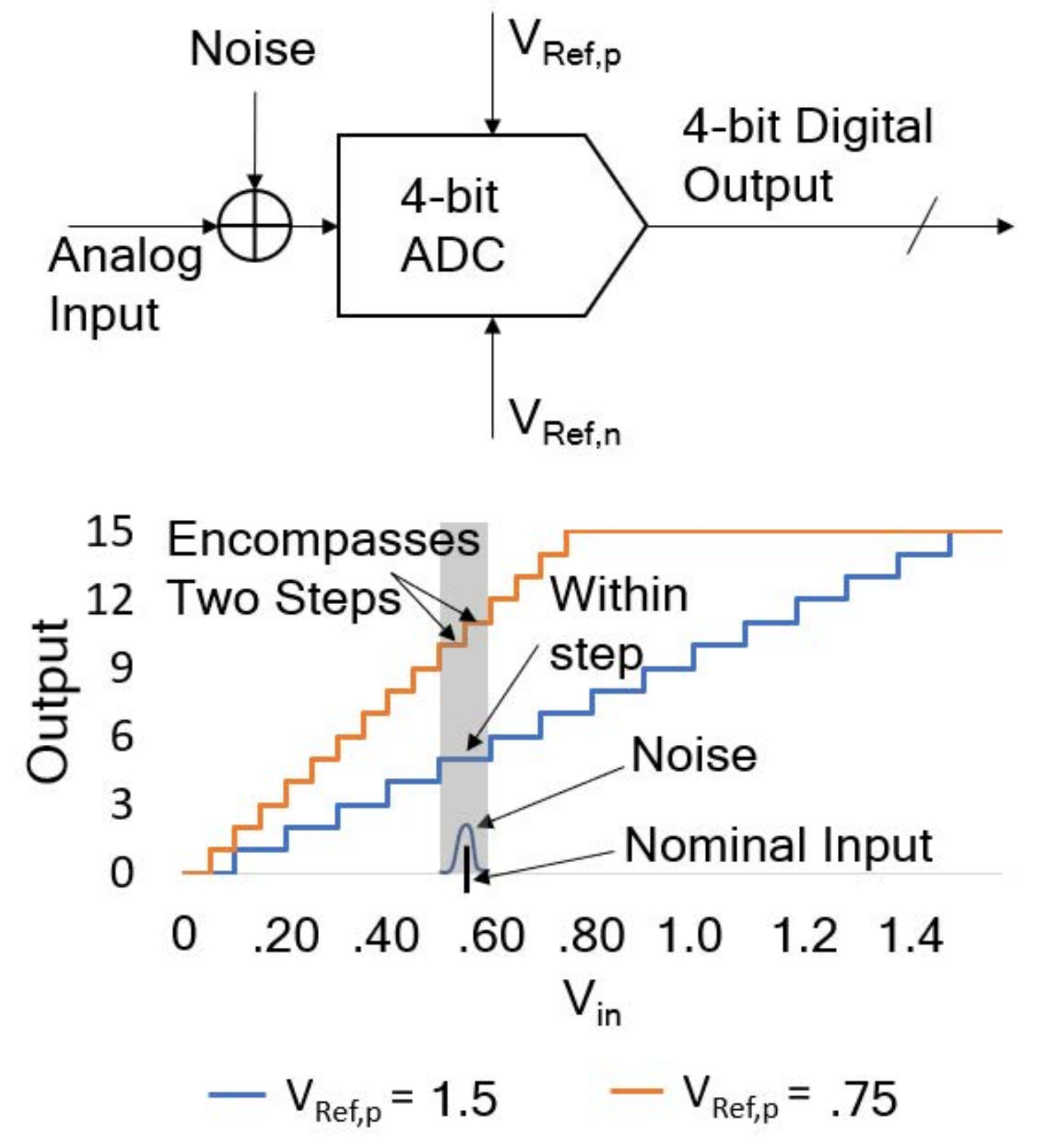}
\caption{Abstract model of an ADC considering the effects of $V_{Ref,p}$ and $V_{Ref,n}$ adjustment as well as the effects of analog noise.}
\label{fig:adc_noise}
\end{figure}

The sparsity described above causes the IMC column signals inputted to the ADC to occupy a reduced range. Exploiting this to mitigate IMC dynamic-range limitations requires adjusting the full-swing range expected by the 8-bit ADC, such that the quantization steps between its 256 levels are reduced. For illustration, Figure \ref{fig:adc_noise} shows an abstract model of a 4-bit ADC. In addition to the analog input and digital output, the ADC has $V_{Ref,p}$ and $V_{Ref,n}$ terminals, which define the upper and lower extremes of the analog input range to be converted (i.e., the ADC quantization steps adjust to uniformly fit between $V_{Ref,p}-V_{Ref,n}$). While this suggests that reducing $V_{Ref,p}-V_{Ref,n}$ benefits quantization noise, and thus ADC resolution, practical ADCs also suffer from internal analog noise effects, which can be modeled as additive input noise. When $V_{Ref,p}-V_{Ref,n}$ is reduced to the point where the quantization steps are comparable to this noise, ADC resolution no longer benefits. While the IMC prototypes considered in this work \cite{65nm, 16nm} exhibit noise well below the quantization steps, such noise will ultimately be of concern with aggressive $V_{Ref,p}-V_{Ref,n}$ reduction.  

Another important consideration around $V_{Ref,p}$/$V_{Ref,n}$ adjustment is the overhead of providing different values to each of the ADCs in a scaled-up IMC architecture. In this work, we consider two cases: (1) the case where the values can be optimally selected for individual ADCs; (2) the more practical case were a few discrete values are provided from which ADCs can select the optimal ones based on the required conversion range.

\section{Evaluation and simulation methodology}
\label{sec:eval}

Simulations were conducted in TensorFlow on GPUs, by implementing custom libraries in order to properly simulate IMC sources of quantization noise during training. The simulations do not model other analog noise effects, which are found to be below the quantization effects of the silicon-measured IMC hardware \cite{65nm, 16nm}. While not considered during the simulations, the possible effects of analog noise will be discussed in Section \ref{conc}.

We chose the CIFAR-10 computer-vision task for analyzing training accuracy as it has been the primary test application for analyzing IMC inferencing accuracy thus far. A 9-layer VGG-lite DNN model was primarily used for testing. Details of the model can be seen in Table \ref{tab:model}. The model is similar to the one used for testing IMC in \cite{65nm} and leads to high utilization of IMC rows across all layers. All convolutional layers use 1$\times$1 strides, a 3$\times$3 filter window, and same padding. Max pooling layers have a 2$\times$2 pooling window. Using a baseline FP32 activations, weights, and gradients, this model achieves a $92.87 \%$ test accuracy after 200 epochs. Tests were also run on a ResNet-18 model as described in \cite{resnet}. For each stack with a stride of 2, the first convolutional layer of the first block has a 2$\times$2 stride for reducing the output size by half. Each block has a residual layer which forwards the block inputs and adds it to the block output. For blocks with a stride of 2, the residual layer is a 1$\times$1 convolutional filter with a stride of 2. All other convolutional layers use 1$\times$1 strides and a 3$\times$3 filter window. Each convolutional layer uses same padding. ResNet-18 uses one average pooling layer with a 4$\times$4 pooling window. The baseline ResNet-18 model achieves $93.15 \%$ test accuracy after 200 epochs. All IMC simulations assume the first, last, and all residual layers are handled by a digital accelerator capable of performing FP32 operations.Keeping the first, last, and residual layers in FP32 is a common approach for many quantized training methods \cite{rad4,8bit,hawq,mixed}. Furthermore, due to the small number of filters these layers are often less efficient to map to IMC and represent a negligible amount of the total number of training MVM operations.

\begin{table}[h]
\caption{DNN Model Details}
\centering
\begin{tabular}{l c c}
\hline
\multicolumn{3}{c}{VGG-lite}\\
\hline \hline
L1: & 128 Conv $\to$ Batch Norm $\to$ ReLU & \\
L2: & 128 Conv $\to$ Max Pool $\to$ Batch Norm $\to$ ReLU & \\
L3: & 256 Conv $\to$ Batch Norm $\to$ ReLU & \\
L4: & 256 Conv $\to$ Max Pool $\to$ Batch Norm $\to$ ReLU &\\
L5: & 256 Conv $\to$ Batch Norm $\to$ ReLU &\\
L6: & 256 Conv $\to$ Max Pool $\to$  Batch Norm $\to$ ReLU &\\
L7: & 1024 Dense $\to$ Batch Norm $\to$ ReLu &\\
L8: & 1024 Dense $\to$ Batch Norm $\to$ ReLU &\\
L9: & 10 Dense $\to$ Batch Norm &\\
[3ex]
\end{tabular}
\begin{tabular}{l r c l}
\hline
\multicolumn{4}{c}{ResNet-18}\\
\hline \hline
L1: & & 64 Conv $\to$ Batch Norm $\to$ ReLU &\\
\multirow{2}{*}{Stack 1:} & \multirow{2}{*}{\bigg( } & 64 Conv $\to$ Batch Norm $\to$ ReLU & \multirow{2}{*}{\bigg) x2} \\
& & 64 Conv $\to$ Batch Norm $\to$ ReLU & \\ 
& & Stride 2 & \\ 
\multirow{2}{*}{Stack 2:} & \multirow{2}{*}{\bigg( } & 128 Conv $\to$ Batch Norm $\to$ ReLU & \multirow{2}{*}{\bigg) x2} \\
& & 128 Conv $\to$ Batch Norm $\to$ ReLU & \\ 
& & Stride 2 & \\ 
\multirow{2}{*}{Stack 3:} & \multirow{2}{*}{\bigg( } & 256 Conv $\to$ Batch Norm $\to$ ReLU & \multirow{2}{*}{\bigg) x2} \\
& & 256 Conv $\to$ Batch Norm $\to$ ReLU & \\ 
& & Stride 2 & \\ 
\multirow{2}{*}{Stack 4:} & \multirow{2}{*}{\bigg( } & 512 Conv $\to$ Batch Norm $\to$ ReLU & \multirow{2}{*}{\bigg) x2} \\
& & 512 Conv $\to$ Batch Norm $\to$ ReLU & \\ 
L18: & & Average Pool $\to$ 10 Dense\\
\end{tabular}
\label{tab:model}
\end{table}

\subsection{Input Quantization Methods}
\label{subsec:input_quant}

Input quantization is required for converting high-precision floating-point values into lower precision values appropriate for IMC. Input quantization is handled in two parts: (1) the values are quantized to their respective integer or radix-4 values; (2) the quantized values are converted to binary representations which can be processed in IMC. For the first part, all MVMs use the quantization methods outlined in \cite{rad4}, which have been modified for our purposes. The following quantization approaches are used for all tests outside the FP32 baseline. Pre-activations use a special version of Parameterized Activation Clipping (PACT), which converts the inputs into unsigned integer activations that have a range of $a_q \in [\, 0,16]\,$. Weights are quantized to symmetric integer values in the range $w_q \in [\,-8,8 ]\,$ using Statistic Aware Weight Binning (SAWB). Gradients are scaled using the gradscale approach and quantized to radix-4 format as presented in \cite{rad4}.

For the second part, all values are converted into a binary representation suitable for IMC bit-serial processing. Activations and weights are converted to the +/-1-binary format according to Equation \ref{eqn:xnor_bin}. Due to the symmetric nature of Equation \ref{eqn:xnor_bin}, activations become a 6-bit binary number with range $[\, -16, 16 ]\,$ and the weights take on a 5-bit binary number. Radix-4 gradients are converted to an 8 bit vector according to Equation \ref{eqn:rad_bin}.  

\subsection{IMC Quantization due to ADC}
\label{subsec:adc_quant}

Beyond input quantization, IMC introduces a second source of quantization within BPBS computation due to the ADC. This ADC quantization is the primary source of noise for IMC and is controlled by the ADC bit precision. Due to ADC size and energy considerations outlined in \cite{65nm}, 8-bit SAR ADC operations are assumed in this work, which provide 256 quantization levels uniformly distributed over an input range set by [$V_{Ref,n}$, $V_{Ref,p}$]. The ADC operation can thus be represented by truncating the input to the nearest 8-bit integer value in this range. That quantized value can then be scaled and offset, as appropriate. Note, with this approach, if the IMC column dynamic range is restricted to 256 values (e.g., by reducing the number of active rows), the ADC will not introduce any quantization. However, if the IMC column dynamic range is greater than 256, the ADC will introduce further quantization noise of the form: $x_q = \lfloor \frac{x \cdot 255}{V_{Ref,p}-V_{Ref,n}} \rfloor $ when x is between $V_{Ref,p}$ and $V_{Ref,n}$. If x is outside this range, $x_q$ is clipped to between 0 and 255 and will take on additional quantization noise. In this work, we assume IMC hardware with 2304 rows, defining the maximum number of active rows. MVMs with larger inner dimensionality are handled by tiled computation across IMC operations.

Due to sparsity of the gradients introduced by the proposed 1-hot encoded exponent representation outlined in Section \ref{subsec:rad4_mapping}, ADC quantization noise can thus be substantially mitigated by appropriately setting the $V_{Ref,p/n}$ values. Three modes of setting $V_{Ref,p/n}$ are explored in this paper: (1) fixed $V_{Ref,p/n}$; (2) variable $V_{Ref, p/n}$; and (3) dual $V_{Ref,p/n}$. Fixed $V_{Ref,p/n}$ provides only a single $V_{Ref,p}$ value and disregards any benefit which may be gained from the sparsity of the input vector. The sparsity of the input vector is still considered after the ADC to ensure proper offsetting of the output. Under variable $V_{Ref,p/n}$, we assume that for every input vector, the nonzero values of the input vector can be summed to determine $V_{Ref,p/n}$ that corresponds to the column output voltage of 2304 to 255 elements. The $V_{Ref,p}$ is not reduced below $V_{prec} = V_{Ref,pmax} \cdot \frac{255}{2304}$, since this value guarantees no ADC quantization noise under our assumptions. Dual $V_{Ref,p/n}$ acknowledges the practical challenges of adjusting all ADC $V_{Ref,p/n}$ at the input-vector rate, and takes the more amenable circuit approach of providing a high-precision $V_{Ref,p}$ value of $V_{prec}$ and a high-input-range $V_{Ref,p}$ value between $V_{Ref,pmax}$ and $V_{prec}$. The high precision $V_{Ref,p}$ was fixed at $V_{prec}$ due to the susceptibility of the gradient MSB to noise. Under all modes $V_{Ref,n}$ is kept at 0V.

\subsection{Energy Models}
\label{subsec:energy}

Energy consumption of the methods presented in this paper is modeled on a T4 GPU and a scaled-up 16nm IMC test chip \cite{16nm}. The analysis considers only the energy of MVM operations, since they dominate DNNs and since other aspects are highly dependent on low-level, often undisclosed, architectural details. The number of multiply-accumulate (MAC) operations for the model given in Table \ref{tab:model} under a batch size of 128 images was calculated for the forward, backward, and weight update MVMs of each layer. Since a MAC operation is composed of two operations (multiply, accumulate), these values were multiplied by 2 to determine the total number of operations. The total operations were then divided by the reported Ops/Watts to determine the J/training iteration.

\begin{table}[h]
\caption{Energy Measurements for 16nm IMC Chip}
\centering
\begin{tabular}{l c}
\hline \hline
\multicolumn{2}{c}{16nm IMC Accelerator \cite{16nm}}\\[0.5ex]
\hline
 Voltage & 0.8 V \\[0.5ex]
 Bit-Cell Multiplication & 0.734 fJ  \\[0.5ex]
 ADC Sample & 346 fJ \\ [0.5ex]
 NMC Data path per output & 243 fJ \\ [0.5ex]
Input Reshape per 32 bit data  & 14.9 fJ \\ [0.5ex]
CIMA Load per 32 bit data & 7360 fJ \\ 
[0.5ex]
\hline
\end{tabular}
\label{tab:energy}
\end{table}

IMC energy consumption is dominated by bit-cell multiplication and ADC accumulation. Bit-cell multiplications are equivalent to the number of MACs per MVM and scale in a similar manner to the GPU energy analysis. ADC accumulation, however, is driven by the column usage as determined by the shape of the inputs which vary for each layer and MVM. Various other sources of energy such as the CIMA load and input reshape were analyzed for each MVM but generally play a secondary effect in energy consumption. All IMC energy consumption factors were calculated based off silicon measurements from \cite{16nm}. A summary of the energy factors for the 16nm IMC chip can be found in Table \ref{tab:energy}.

Beyond energy efficiency, latency and throughput are important measurements of performance. The one-hot encoding of radix-4 gradients actually reduces the number of CIMA cycles per gradient to 7 from 8 for a more traditional 8-bit integer approach. Due to the incredibly low bit-precision of the gradients, the one-hot encoding doesn't even double the number of CIMA cycles for a theoretical 4 bit floating point implementation. As mentioned in Section \ref{subsec:dyn_range_imc}, constant switching of the ADC reference voltages may significantly degrade latency and throughput; however, their measurement is outside the scope of this work.

\section{Analysis and Results}

\subsection{Variable $V_{Ref,p/n}$}
Initial tests were conducted to determine if the proposed 1-hot-encoded radix-4 +/-1-binary format and ultra-low 4-bit training techniques could be used to accurately implement DNN training on IMC accelerators. Four tests on two different models were conducted with the IMC simulator using variable $V_{Ref,p/n}$ mode. In three of these tests, the IMC simulator was used for only one of the training MVMs. In the fourth test, the IMC simulator was used for all three training MVMs. The results are presented in Table \ref{tab:varVRef}. For the VGG-lite model all test accuracies are within $1\%$ of the baseline max test accuracy for the given DNN model.

The more complex ResNet-18 model shows some degradation, but still high test accuracy results. Forward and weight-update MVM tests achieve within $3\%$ of the baseline test accuracy. The preliminary test show some degradation in the backward MVM for the ResNet model, within $5\%$ of the baseline (with the first two of four ResNet-18 stacks mapped, representing almost half of the MACs). The ResNet-18 model does not perform as well as the shallower VGG-lite model when IMC computations are introduced. While the 1-hot encoding technique reduces the ADC quantization noise, it does not entirely eliminate it. Since IMC techniques add ADC quantization noise at every layer IMC is implemented in, the accuracy for deeper networks can tend to decrease as the number of layers increases. This presents an important area for further research exploring methods of increasing noise tolerance for very deep neurla networks. Some promising examples are \cite{noiseaware,noisetol,contrastive}.

\begin{table}[h]
\caption{Effect of IMC on Training MVMs using Variable $V_{Ref,p/n}$}
\centering
\begin{tabular}{l c r}
\hline \hline
Model & IMC Use & Max Test Accuracy \\ [0.5ex]
\hline
VGG-lite & Baseline & $92.87 \%$ \\
& Forward MVM & $92.02 \%$ \\
& Backward MVM &  $92.47 \%$ \\
& Weight Update MVM & $91.97 \% $ \\
& All MVM & $92.30 \% $ \\ [0.5ex]
\hline
ResNet-18 & Baseline & $93.15 \%$ \\
& Forward MVM & $91.51 \%$ \\
& Backward MVM$^*$ &  $88.57 \%$ \\
& Weight Update MVM & $91.58 \% $ \\
& All MVM$^*$ & $88.73 \% $ \\
[1ex]
\hline
\multicolumn{3}{l}{* Only the first two ResNet stacks are mapped to IMC}\\
\multicolumn{3}{l}{for the Backward MVM}\\
\end{tabular}
\label{tab:varVRef}
\end{table}

As stated in Section \ref{sec:eval}, the simulations only factor in noise due to input and ADC quantization. Further degradation could be realized for actual IMC training because of the effects outlined in Section \ref{subsec:dyn_range_imc}. Furthermore, as previously mentioned, variable $V_{Ref,p/n}$ introduces challenges at the circuit-implementation level.

\subsection{Fixed $V_{Ref,p/n}$}

\begin{figure}[htb]
\centering
\begin{subfigure}{.5\textwidth}
  \centering
  \includegraphics[width=1\linewidth]{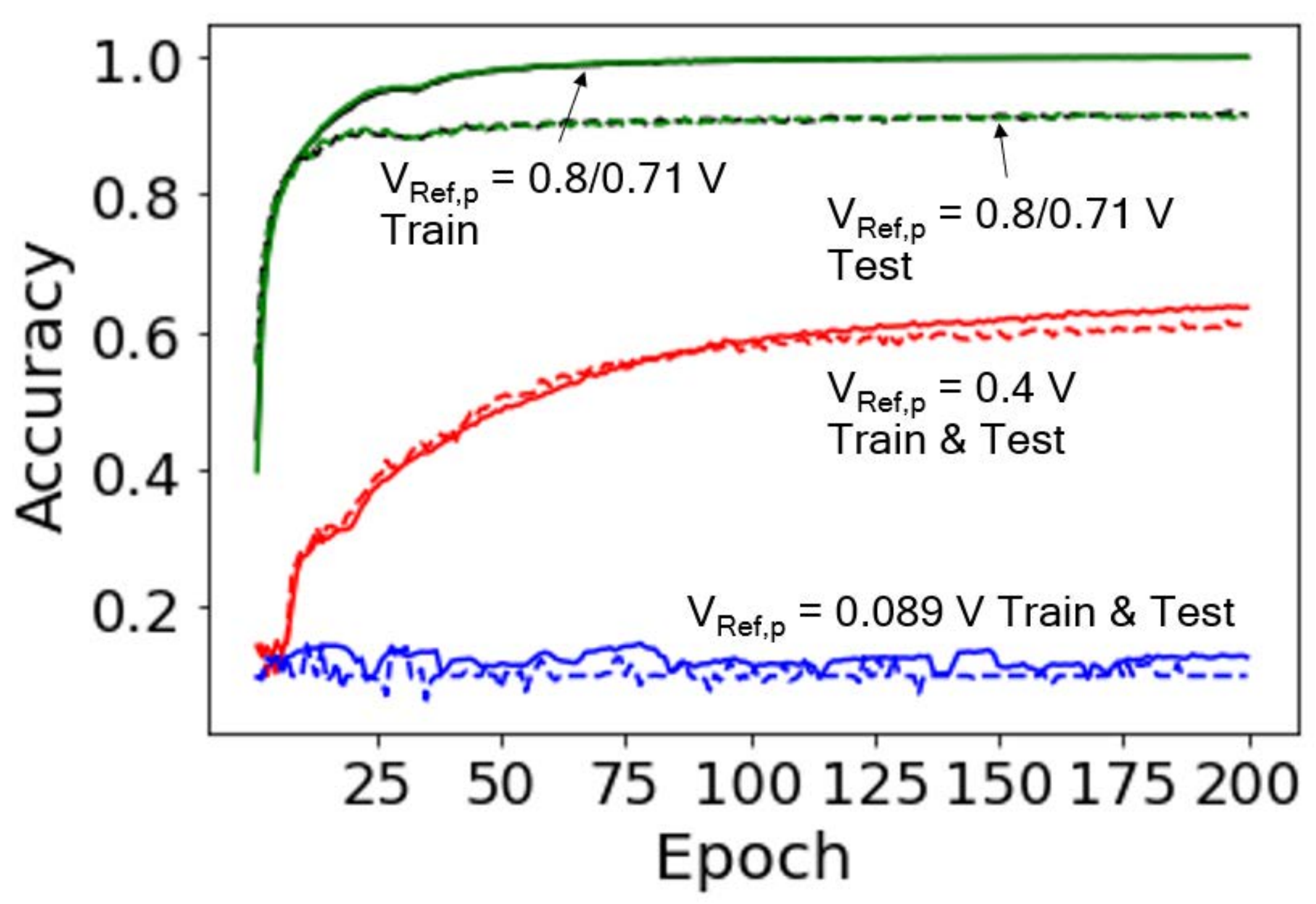}  
  \caption{}
  \label{fig:sub_uni_for}
\end{subfigure}
\begin{subfigure}{.5\textwidth}
  \centering
  \includegraphics[width=1\linewidth]{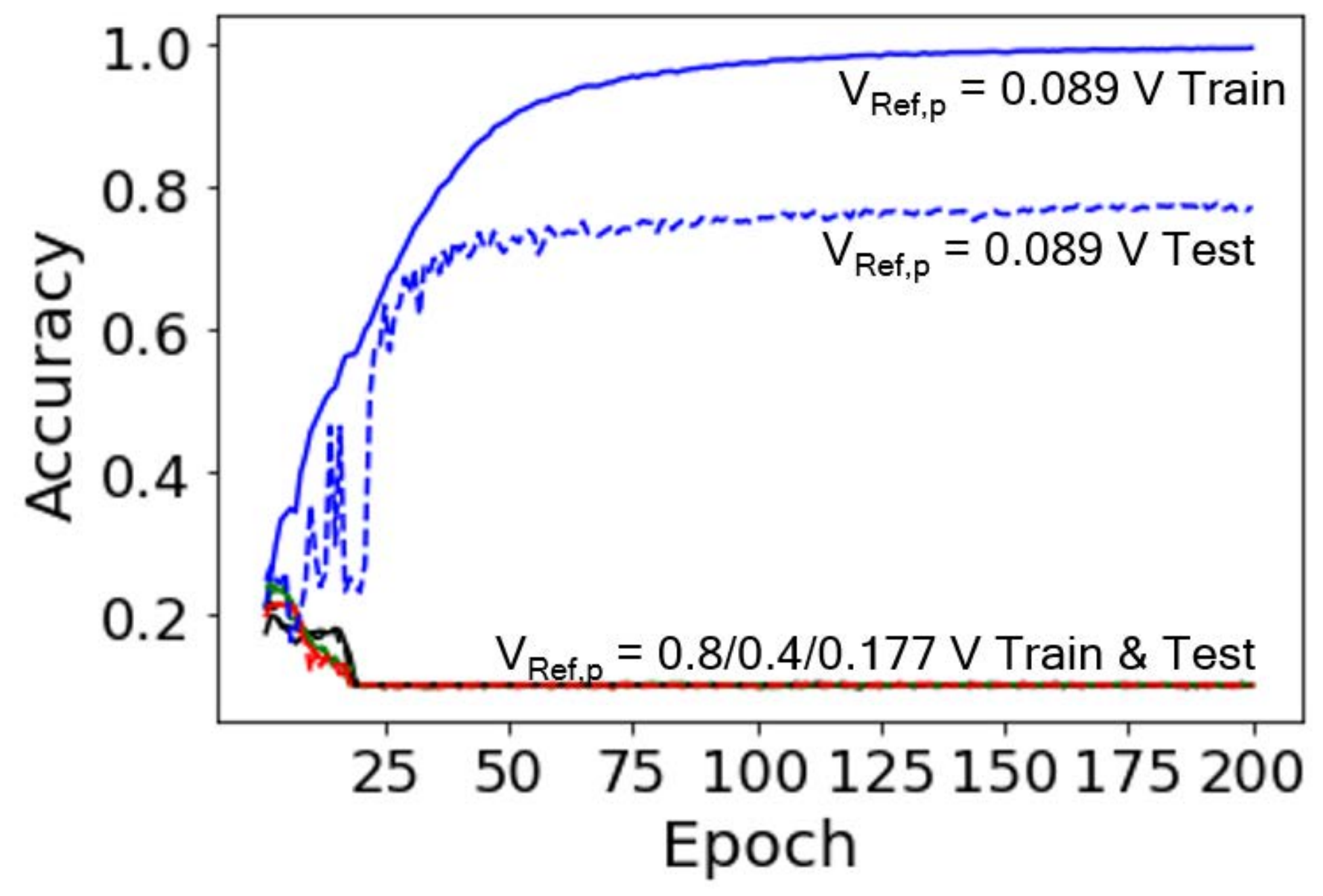}  
  \caption{}
  \label{fig:sub_uni_back}
\end{subfigure}
\caption{Analysis of training MVM accuracies for different values of fixed $V_{Ref,p}$ on the VGG-lite model}, showing: (a) forward MVMs maintain accuracy when using $V_{Ref,p}$'s that enable a large ADC input range (i.e. covering voltages that include $>$ 2000 rows); (b) backward MVMs exhibit low accuracy when using high $V_{Ref,p}$, causing large ADC quantization noise, and exhibit limited generalization when using low $V_{Ref,p}$. 
\label{fig:uni_vref}
\end{figure}

Additional tests were run on the VGG-lite model. Keeping the $V_{Ref,p/n}$ fixed significantly eases circuit implementation, and has already been shown to enable high inferencing accuracy when $V_{Ref,p}$ is kept high to match the maximum range of the ADC input \cite{65nm, 16nm}. Figure \ref{fig:sub_uni_for} shows that operating IMC with a high constant $V_{Ref,p}$ for the forward MVM provides for high training and testing accuracy. However, reducing the $V_{Ref,p}$ substantially causes both training and testing accuracy to degrade due to the saturation of the ADC, which causes clipping when $V_{Ref,p}$ is exceeded by the actual ADC input.

Conversely, for the backward MVM, a fixed $V_{Ref,p/n}$ approach has significant drawbacks. In order to capture the high precision and small dynamic range of the gradient input vectors, the $V_{Ref,p}$ can be set low. This setting performs well for training accuracy, but exhibits significant degradation for testing accuracy, as can be seen in Figure \ref{fig:sub_uni_back}. Further, when $V_{Ref,p}$ is increased and the precision degraded, the backward MVM drastically degrades in performance. While the low $V_{Ref,p}$ enables training with the backward MVM, there does not appear to be a single fixed $V_{Ref,p}$ that provides baseline-level training generalization (i.e., testing accuracy).

Unfortunately, a single fixed $V_{Ref,p}$ does not exist if we wish to use IMC for multiple training MVMs. Either a high $V_{Ref,p}$ must be supplied to ensure adequate forward-MVM accuracy or a low $V_{Ref,p}$ must be supplied to provide adequate backward-MVM accuracy. However, these tests do suggest that both the forward and backward MVMs may be implemented with a limited number of different $V_{Ref,p/n}$ values.

\subsection{Dual $V_{Ref,p/n}$}

\begin{figure*}[htb]
\centering
\begin{subfigure}{.475\textwidth}
  \centering
  \includegraphics[width=1\linewidth]{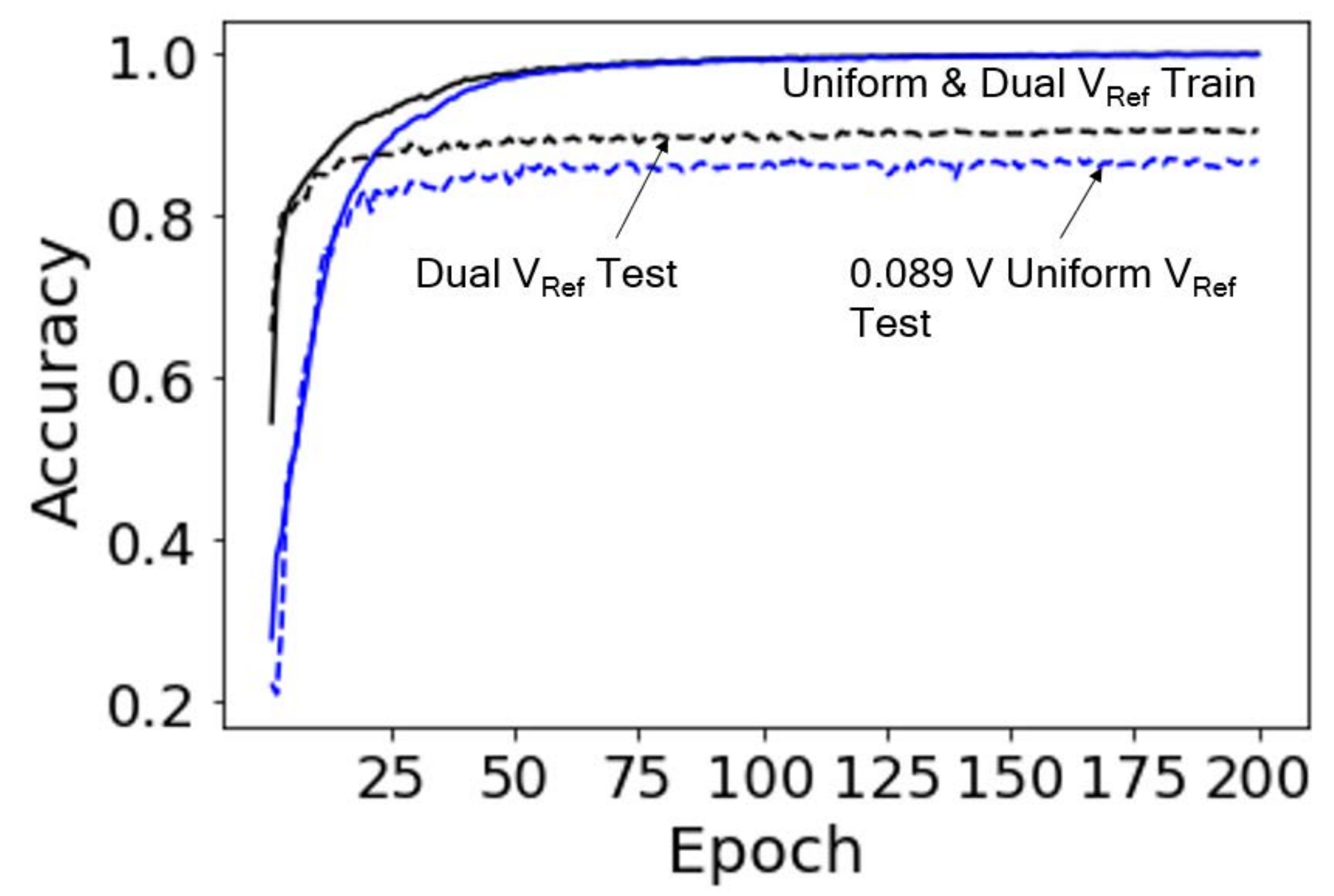}  
  \caption{}
  \label{fig:sub_dual_back}
\end{subfigure}
\hfill
\begin{subfigure}{.475\textwidth}
  \centering
  \includegraphics[width=1\linewidth]{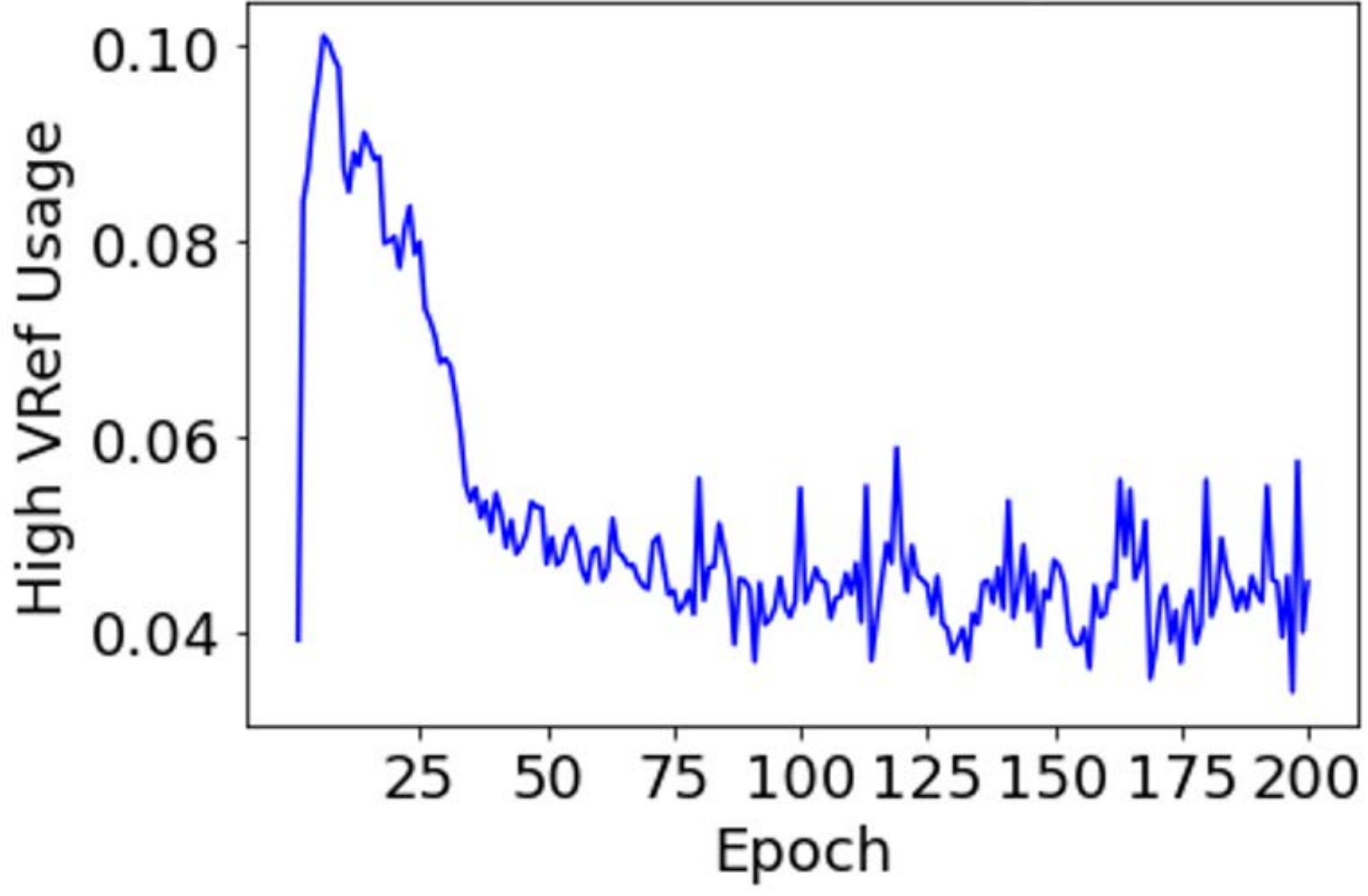}  
  \caption{}
  \label{fig:sub_vref}
\end{subfigure}
\vskip\baselineskip
\begin{subfigure}{.475\textwidth}
  \centering
  \includegraphics[width=1\linewidth]{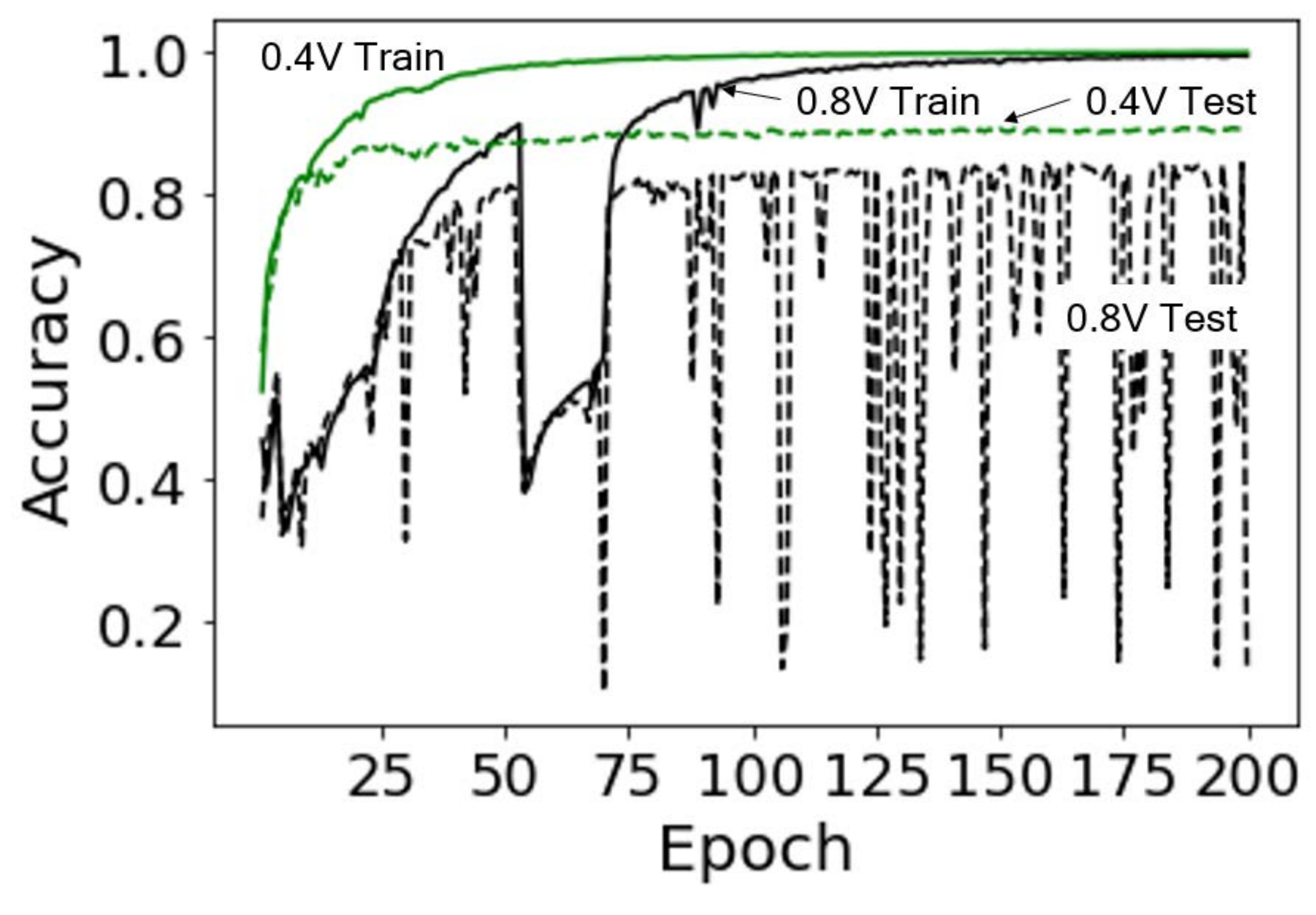}  
  \caption{}
  \label{fig:sub_dual_wu}
\end{subfigure}
\hfill
\begin{subfigure}{.475\textwidth}
  \centering
  \includegraphics[width=1\linewidth]{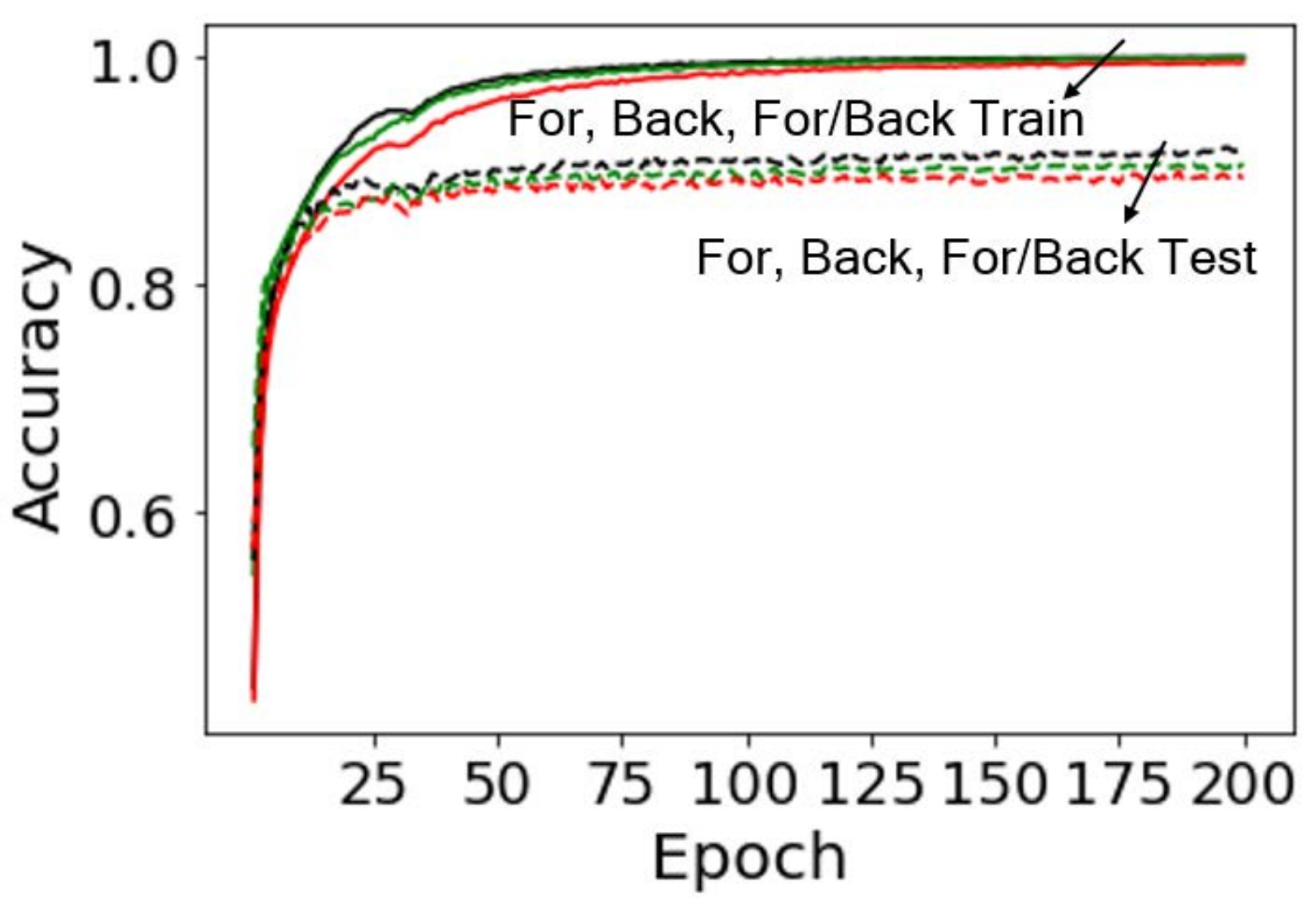}  
  \caption{}
  \label{fig:sub_dual_for_back}
\end{subfigure}
\caption{Dual $V_{Ref,p/n}$ recovers generalization of the backward MVM, but still exhibits degradation in weight-update MVM accuracy, as shown by: (a) implementing two $V_{Ref,p}$'s in the backward pass recovers the test accuracy performance as compared to the fixed $V_{Ref,p}$; (b) Under dual $V_{Ref,p/n}$, high $V_{Ref,p}$ usage for backward MVMs is limited to between 4 and 10\% of the total MVMs; (c) even with dual $V_{Ref,p/n}$, the weight-update MVM leads to degradation in testing accuracy; (d) dual $V_{Ref,p/n}$ for forward and backward MVMs results in high training accuracy with minimal degradation in test accuracy.}
\label{fig:dual_vref}
\end{figure*}

Dual $V_{Ref,p/n}$ was explored where a low $V_{Ref,p}$ of 0.089 V was used, and a number of high  $V_{Ref,p}$'s were selected based on varying ADC quantization noise and clipping of high input values. Ultimately, a high $V_{Ref,p}$ of 0.8 V was selected since this high $V_{Ref,p}$, with the low $V_{Ref,p}$, performed the best for the forward and backward MVMs respectively, under the fixed $V_{Ref,p/n}$ mode. As seen in Figure \ref{fig:sub_dual_back}, allowing high and low $V_{Ref,p}$'s results in a recovery of generalization performance for the backward MVM.

While the forward MVM would only require the high $V_{Ref,p}$, maintaining the improved test accuracy for the backward MVM would require $V_{Ref,p}$ switching. Figure \ref{fig:sub_vref} shows that use of the high $V_{Ref,p}$ for the backward MVM is limited, applying to less than 10\% of the total backward MVMs. High $V_{Ref,p}$ usage also tends to occur more frequently in the earlier epochs with a significant decrease in high $V_{Ref,p}$ usage by the $50^{th}$ epoch. This shows that most of the $V_{Ref,p}$ switching would occur with predominately high $V_{Ref,p}$ for forward MVM operations and predominantly low $V_{Ref,p}$ for backward MVM operations.

In the case of the weight-update MVM, the two $V_{Ref,p}$'s continue to exhibit degradation, as seen in Figure \ref{fig:sub_dual_wu}. Using a high $V_{Ref,p}$ of 0.8V causes significant divergences in the training accuracy before finally converging to $100 \%$ training accuracy. Furthermore, the test accuracy fails to achieve a stable convergence. Reducing the high $V_{Ref,p}$ to 0.4V eliminates the convergence issues but creates similar generalization issues as the backward MVM with fixed $V_{Ref,p}$. This degradation in the weight update-MVM accuracy suggest that the dual $V_{Ref,p/n}$ approach does not yet fully enable mapping all three of the training MVMs to IMC.

However, using the $V_{Ref,p}$ pair of 0.8V and 0.089 V performs well for both the forward and backward MVMs resulting in a test accuracy of $89.89\%$. Figure \ref{fig:sub_dual_for_back} shows that this mode of operation sees a minimal degradation in training and test accuracy convergence when applied to the forward and backward MVMs independently as well as for the forward and backward MVMs together. This dual $V_{Ref,p/n}$ approach provides a feasible means of leveraging IMC advantages for training. 

\subsection{Energy Analysis}

\begin{figure}[htb]
\centering
\includegraphics[width=1\linewidth]{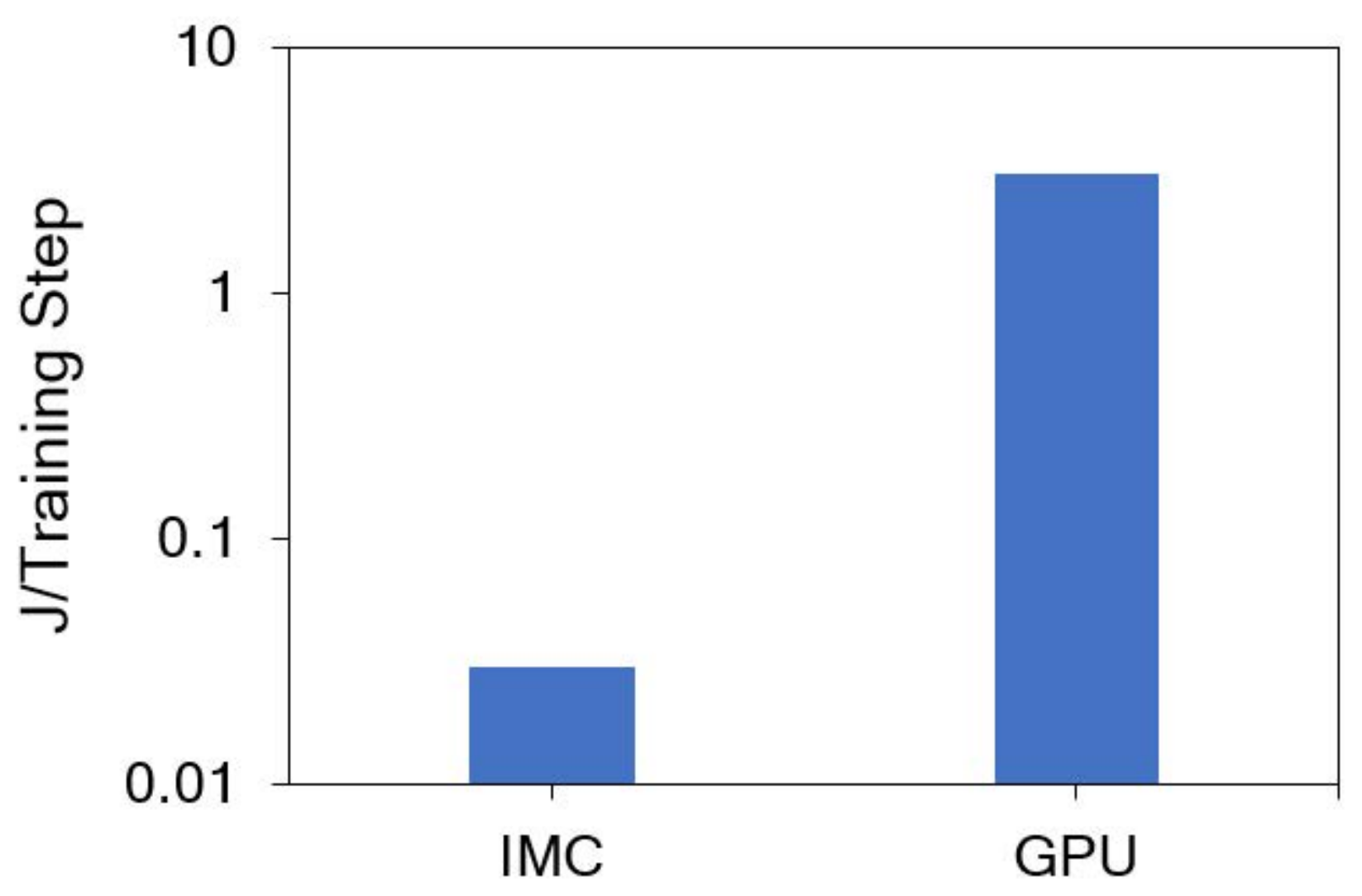}
\caption{While training DNNs generally takes multiple joules of energy per training step, this can be reduced to less than 100 mJ by using IMC accelerators.}
\label{fig:tot_ene}
\end{figure}

Variable and dual $V_{Ref,p/n}$ demonstrate the ability to extend IMC efficiency advantages to training (i.e., beyond just the forward MVM for inferencing). Under variable $V_{Ref,p}$, IMC efficiencies can be extended to all three training MVMs for layers 2-8 of the model considered. This results in over two orders of magnitude energy reduction per training step, as seen in Figure \ref{fig:tot_ene}. Such high overall reduction in energy is only possible because of extension to all three training MVMs, given their equal contribution of MACs. 

Implementing any one of the MVMs on a GPU creates an energy bottleneck. For example, under dual $V_{Ref,p/n}$ only the forward and backward MVMs can be implemented in IMC. This limits the energy reduction to 3$\times$ the GPU baseline. The energy reduction, relative to using dual $V_{Ref,p/n}$ for only the forward MVM, is thus $2 \times $ greater. 

\begin{figure}[!htb]
\centering
\begin{subfigure}{.5\textwidth}
  \centering
  \includegraphics[width=1\linewidth]{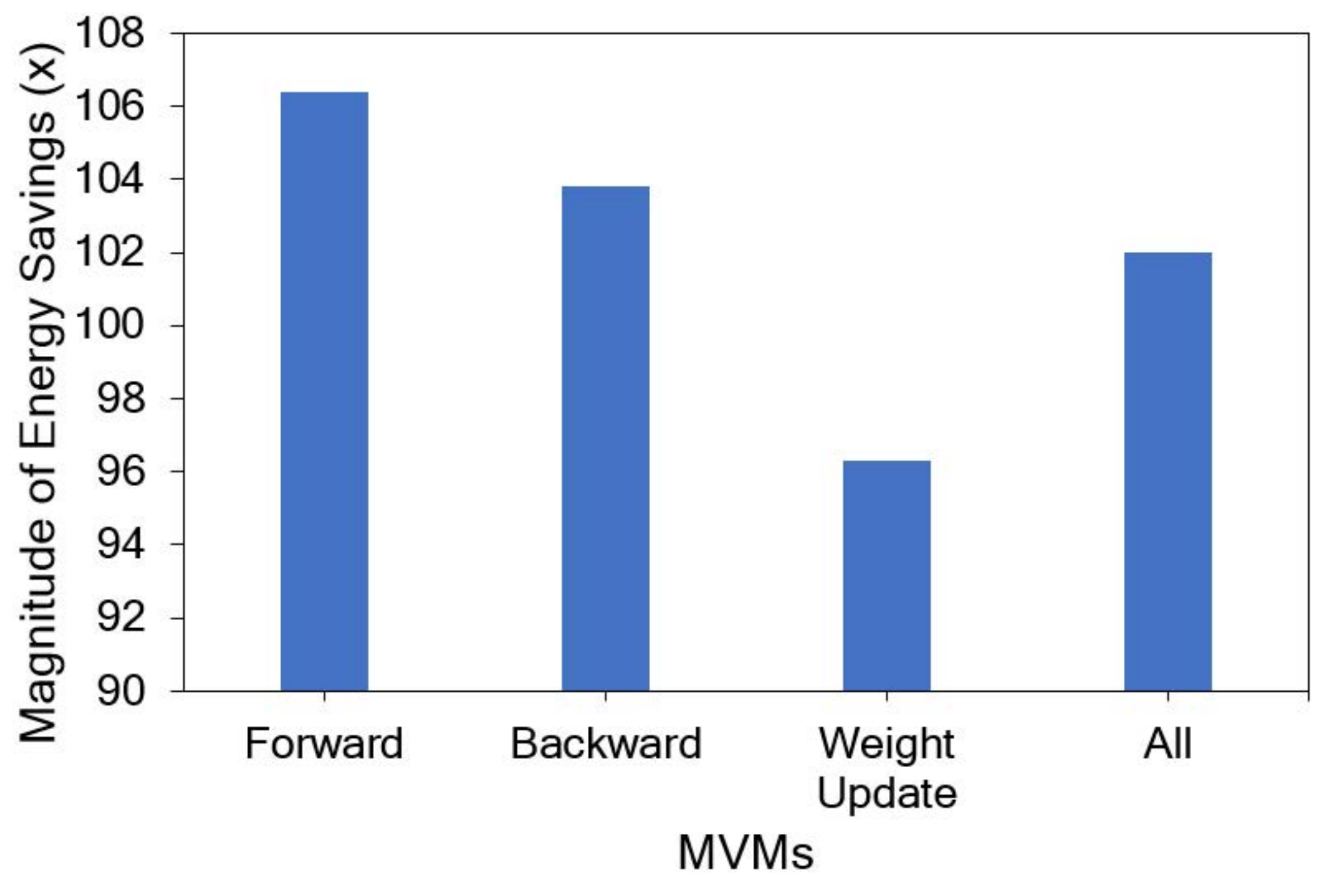}  
  \caption{}
  \label{fig:sub_ene_mvm}
\end{subfigure}
\begin{subfigure}{.5\textwidth}
  \centering
  \includegraphics[width=1\linewidth]{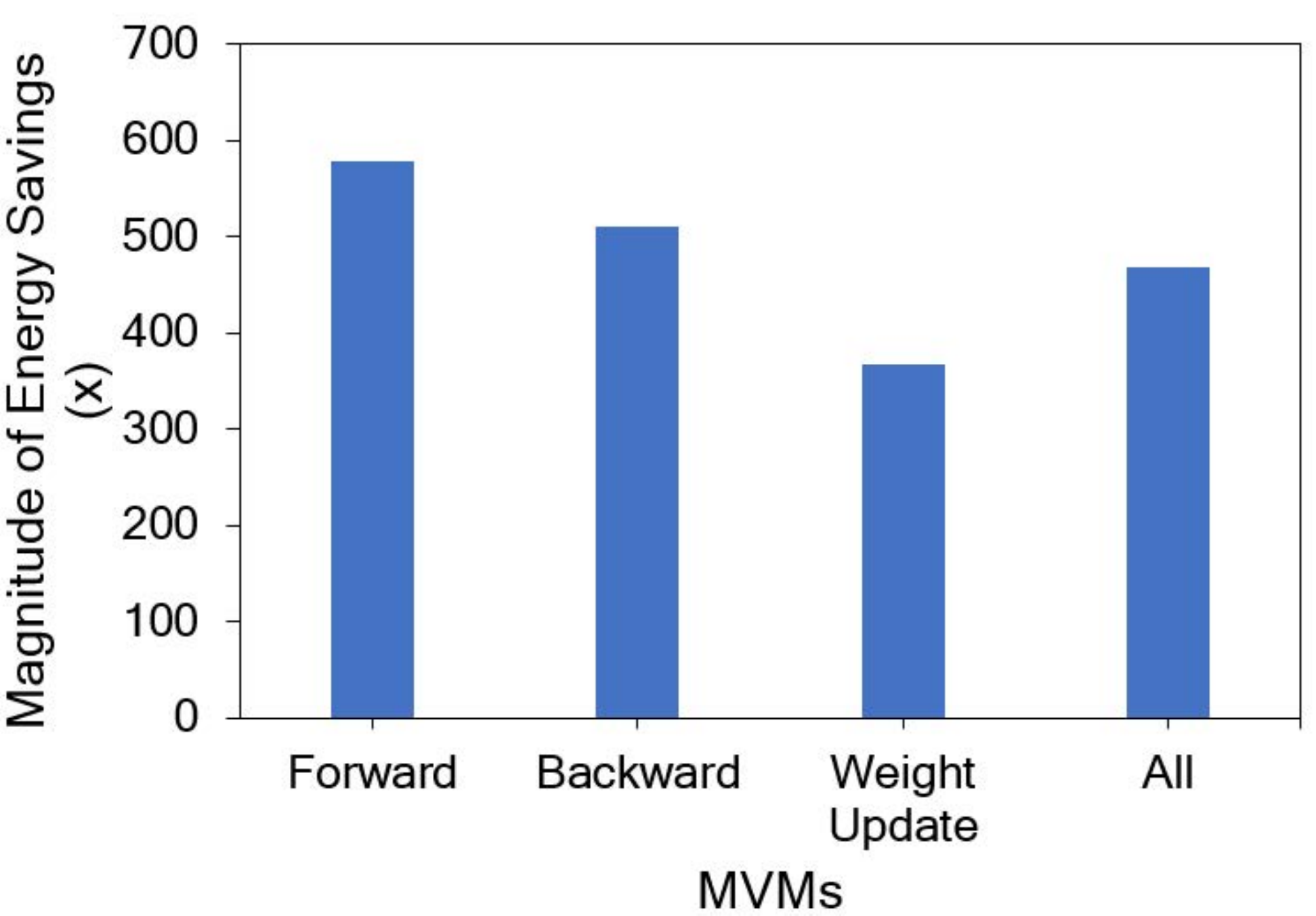}  
  \caption{}
  \label{fig:sub_ene28}
\end{subfigure}
\caption{  Extending IMC to each of the training MVMs greatly reduces energy consumption. a) Under a model where only the first and last layer MVMs are implemented in a GPU, IMC training results in over a 100x saving in energy consumption. b) Comparison between energy consumption of the IMC and GPU on layers 2-8. }
\label{fig:energy}
\end{figure}

Even with variable $V_{Ref,p/n}$, energy savings are still throttled. The input quantization methods used in this paper still require the first and last layer to be kept at FP32 precision operations in order to maintain training accuracy. Fortunately, these two layers together account for less than 1 \% of the training MVM operations. Removing these two layers from the analysis results in over a 5$\times$ further energy decrease as can be seen in Figure \ref{fig:energy}. Figure \ref{fig:sub_ene28} permits a direct comparison between the energy consumption of IMC versus the energy consumption of GPUs. The energy consumption savings by using IMC are so drastic that the use of GPUs on less than 1 \% of the training MVMs results in a bottleneck for reducing training energy.  

\section{Discussion \& Conclusion}
\label{conc}
Adapting IMC to training has previously been elusive due to the higher gradient precision necessary for training-specific MVMs, the large amount of analog noise inherent in previous systems, and the fundamental dynamic range limitations of IMC. Recent advances have closed the first two gaps with quantized training approaches reducing the gradient precision to below 8-bits and capacitor-based IMC reducing analog noise below quantization effects. The presented radix-4 +/-1-binary format gradients builds on these advances by mapping radix-4 gradients to XNOR computations within low analog noise IMC hardware. Furthermore, this format substantially addresses the dynamic range limitations of IMC by using a 1-hot encoding of the exponents which creates high levels of sparsity. These high sparsity levels are exploited to reduce IMC dynamic-range requirements and improve the performance of training-specific MVMs. We have demonstrated that these techniques can achieve high training accuracy, allowing us to apply the benefits of IMC to training in order to reduce the energy required to train DNNs by two orders of magnitude. Our work expands the capability to train DNN's efficiently by reducing ADC quantization without increasing ADC precision. By taking advantage of the sparsity in the 1-hot encoded radix-4 gradients, we can maintain high IMC row parallelism while reducing dynamic-range requirements, thereby allowing us to achieve high energy efficiency as seen in Table \ref{tab:soacomp}.

\begin{table*}[htb]
\caption{Performance Comparison of Training Accelerators}
\centering
\begin{tabular}{l | c | c | c| c| c | c}
\hline \hline
 & GPU \cite{t4}  & \cite{lnpu} & \cite{bruce} & \cite{2way} & \cite{rram} & This Work \\ [0.5ex]
\hline
Technology & 12nm  & 65nm & 14nm & 28nm & 22nm & 16nm \\
Model & Various  & VGG, ResNet, AlexNet & CNN, MLP, LSTM & VGG, ResNet, DenseNet & VGG & VGG, ResNet \\
Dataset & Various  & CIFAR100, ImageNet & - & CIFAR10, ImageNet & CIFAR10 & CIFAR10 \\
Architecture & Digital  & Digital & Digital & SRAM IMC & RRAM IMC & SRAM IMC \\
Input Precision (bits) & 32/16/8/4  & 8/16 & 16 & 8 & 8 & 4--8 \\
Output Precision (bits) & 32/16/8/4  & 8/16 & 16 & 5 & 6 & 8 \\
MAC Parallelism & -  & - & - & 16 & - & 2304 \\
MAC TOPS/W & 0.116  & 3.48 & 15 & 17.2 & 26.66 & 54.2 \\
CIFAR-10 Test Accuracy & 92.87  & 92.87 & 92.87 & 90 & 81 & 92.3 \\
[1ex]
\hline
\end{tabular}
\label{tab:soacomp}
\end{table*}

Beyond demonstrating methods and benefits of applying IMC to training, this work also highlights a number of current limitations and promises for further improvements in IMC training. Reducing the dynamic-range requirements of IMC without clipping requires a large number of $V_{Ref,p}$ values which increases circuit complexity. However, sparsity of the gradients suggests that the dynamic-range requirements of IMC can be met with far fewer $V_{Ref,p}$ values. We show that dual $V_{Ref,p/n}$ mode can maintain high levels of performance in the forward and backward MVMs while reducing the number of $V_{Ref,p}$ values to two. This reduction in $V_{Ref,p}$ values greatly reduces the overhead necessary to implement training in IMC and improves the practicality. A more exhaustive search may reveal sets of $V_{Ref,p/n}$ that optimize either training performance, complexity, energy consumption, or some combination of the three.

Limitations of quantization in the first and last layers pose a bottleneck to increased reductions in energy consumption for training. Despite accounting for less than 1 \% of operations, the first and last layers reduce the energy reduction benefits by 5 $\times$. As maintaining the first and last layer at higher precision is a common technique for quantized training, this bottleneck highlights the need for approaches that can be ubiquitously applied across DNN models. Extending quantization methods to these layers could result in even larger benefits for IMC training.

In summary, we have made three major advancements toward realizing IMC training of DNNs:
\begin{enumerate}
\item Demonstrated a method for mapping radix-4 gradients to a +/-1-binary format with 1-hot encoded exponent. This format enables XNOR computations within IMC hardware, and exploits high levels of sparsity to reduce dynamic-range requirements and improve the performance of training-specific MVMs.
\item Explored the feasibility of IMC training and their practical limitations. While variable $V_{Ref, p/n}$ methods require supplying a large number of $V_{Ref,p}$ values and frequent ADC calibration, limiting the number of $V_{Ref,p}$ levels results in a reduction in performance for the backward and weight update MVMs.
\item Showed the energy benefits of extending IMC to training. Using methods that permit IMC training for the forward, backward, and weight update MVMs would reduce MVM energy consumption by over two orders of magnitude, while limiting IMC training to the forward and backward MVM results in a 3$\times$ reduction. 
\end{enumerate}

Future work includes modeling additional sources of noise during ADC accumulation. These noise sources include capacitor mismatch, charge injection, thermal/shot electronic noise, and other analog non-idealities. While previous IMC work has shown that these sources of noise fall well below quantization noise, many of the training-specific MVMs greatly reduce the $V_{Ref,p}$ value and therefore the quantization step. As presented in Section \ref{subsec:dyn_range_imc}, these sources of noise could have notable effects to the ADC output especially when $V_{Ref,p}-V_{Ref,n}$ is reduced. By further characterizing and modeling IMC noise, we can determine how IMC training will perform when implemented on-chip. Additional techniques for training DNNs under analog noise may thus be required, which is a separate topic from training under practical quantization noise considered in this work.

Our primary motivation revolves around decreasing the compute energy of MVM operations in-memory and we therefore have avoided a larger analysis of the impact of data movement for training with IMC. The impact of data movement is more complex due to its dependency on the execution scheduling and mapping in the presence of hardware parallelism. Exploring system level techniques for optimizing the execution of DNN training on IMC is an important area of future work in order to fully achieve the benefits of training with IMC.

This work serves as a starting point that can also leverage a number of mixed-precision training strategies that could deliver the right mix of precision and energy efficiency. The IMC architecture analyzed in this work allows for the flexibility of input bit-precision which could leverage mixed precision input quantization approaches as outline in \cite{hawq,mixed}. Furthermore, various $V_{Ref,p/n}$ strategies could be implemented to deliver mixed precision output quantization approaches. Beyond flexibility within the IMC architecture, mixing the mapping of IMC and standard-digital computing approaches to specific DNN layers as in \cite{hyper} could also provide new trade-offs between performance and energy efficiency.

The 1-hot encoding approach is similar to methods used by spiking neural networks, such as spike-trains presented in \cite{spiking,datta}. Energy and latency benefits of simple IMC computations are maintained by encoding additional information in the temporal domain, in this case the order of the inputs. Further latency and energy benefits may be realized through more effective use of temporal domain information.

Finally, the presented analysis on VGG-lite for CIFAR-10 classification demonstrates how quantization effects can be managed for IMC training. Current work is investigating a broader range of models, tasks, and memory technologies. Initial results show that quantization effects are well managed with ultimate performance limited by other analog noise sources. This is particularly true for IMC with emerging memory technologies where analog noise sources typically dominate \cite{rram}.  We hope the analysis will provide insights on how to specify analog noise levels for different model accuracy requirements.

\section*{Acknowledgments}
This research is supported through funding by the United States Air Force under the Dr. Heather Wilson STEM PhD Program. The views and conclusions contained herein are those of the authors and should not be interpreted as necessarily representing the official policies or endorsements, either expressed or implied, of the U.S. Air Force or the U.S. Government.

\section{References Section}


\bibliographystyle{IEEEtran}

\newpage

\section{Biography Section}

\begin{IEEEbiography}[{\includegraphics[width=1in,height=1.25in,clip,keepaspectratio]{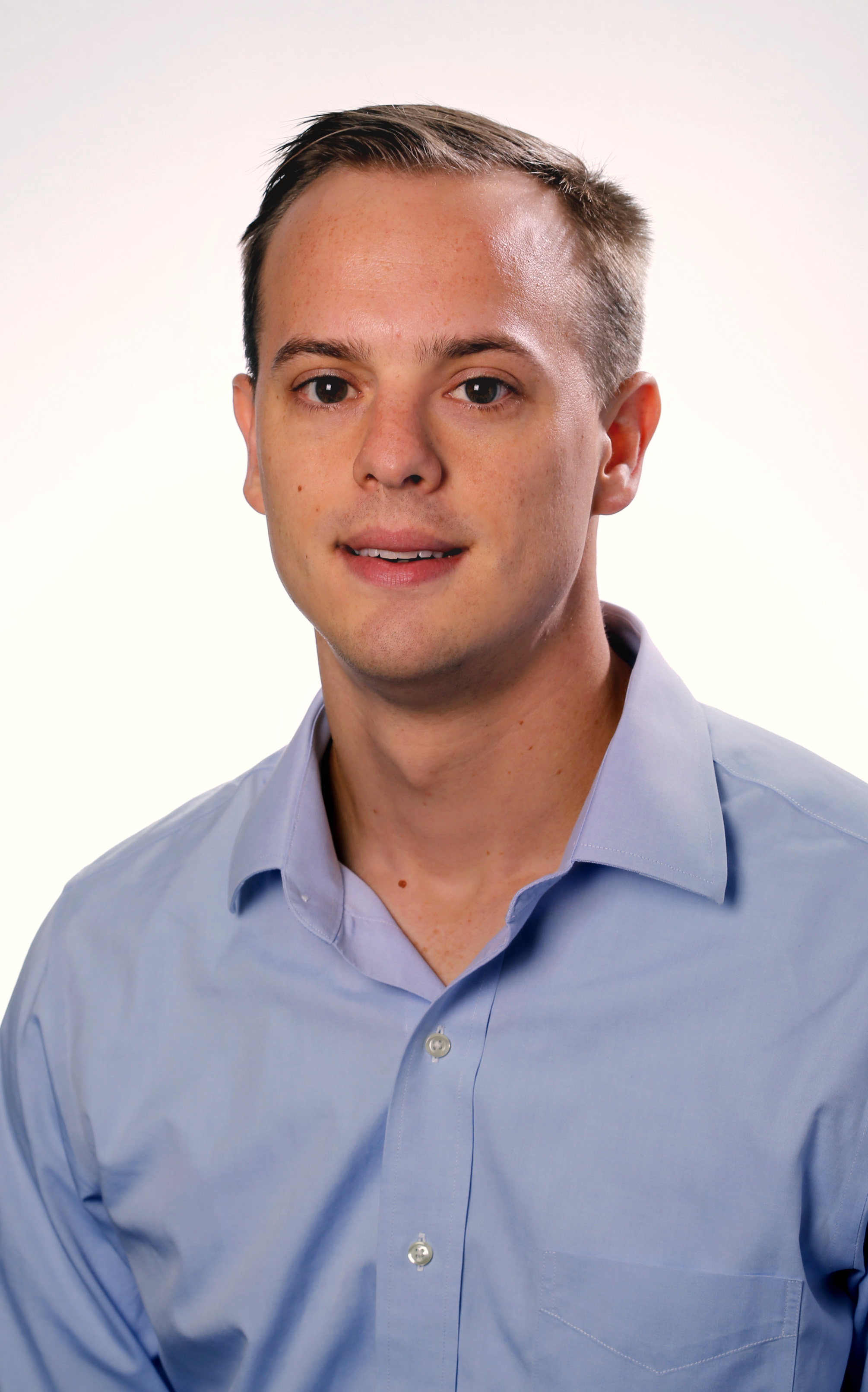}}]{Christopher Grimm}
(Student Member, IEEE) received the B.S. degree in computer engineering from The United States Air Force Academy (USAFA), US Air Force Academy, CO, USA, in 2014, the M.S. degree in Aeronautics and Astronautics from the Massachusetts Institute of Technology (MIT), Cambridge, MA, USA, in 2016, and the M.A. degree in electrical engineering from Princeton University, Princeton, NJ, USA, in 2021, where he is currently pursuing the Ph.D. degree.\\
His research focuses on deep learning training applications for ultra-low-energy systems. His current research interests include mixed-signal computing, computer vision, and noise-tolerant machine learning.
\end{IEEEbiography}
 
\vspace{11pt}

\begin{IEEEbiography}[{\includegraphics[width=1in,height=1.25in,clip,keepaspectratio]{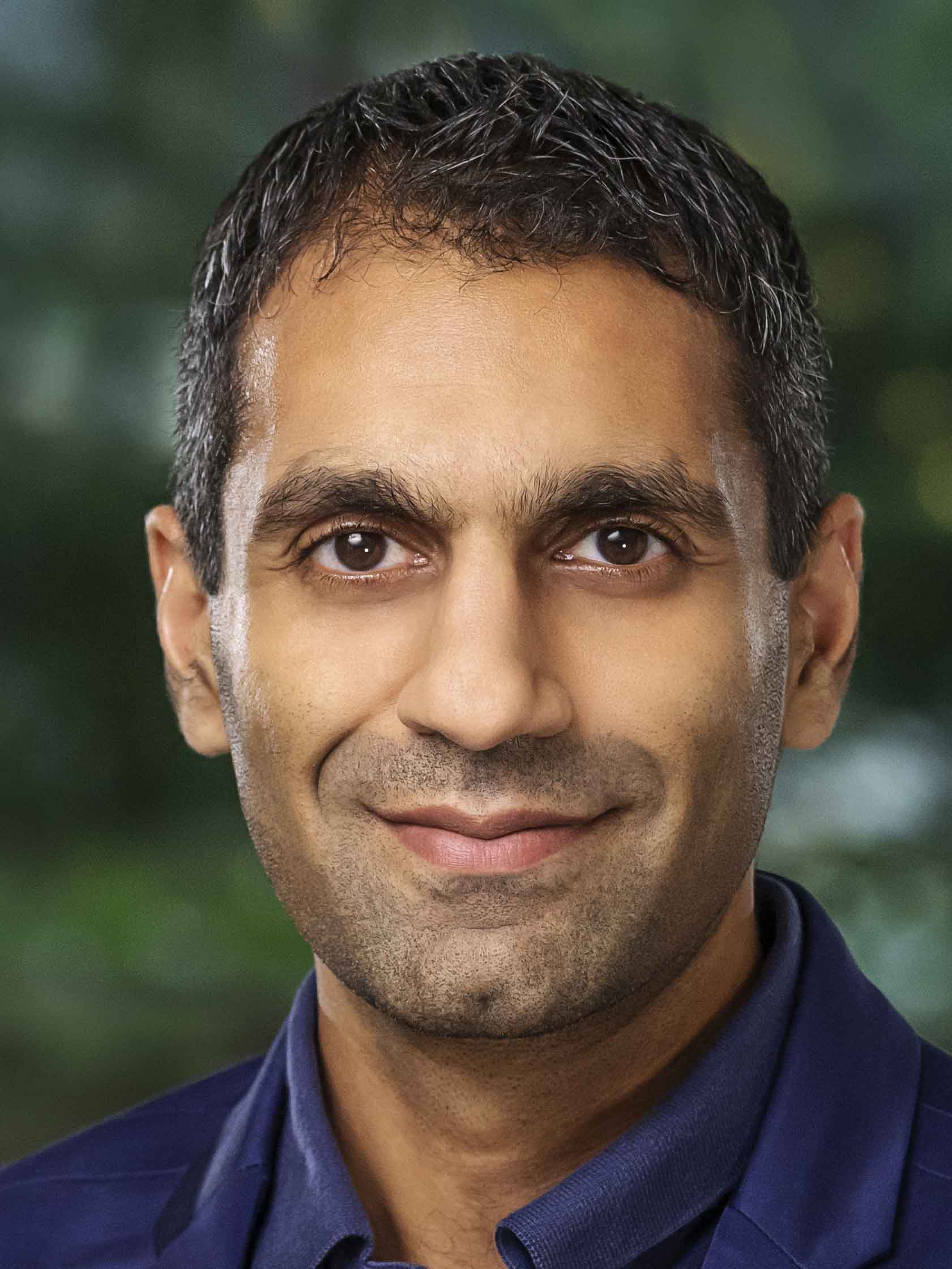}}]{Naveen Verma}
(Member, IEEE) received the B.A.Sc. degree in electrical and computer engineering from The University of British Columbia (UBC), Vancouver, BC, Canada, in 2003, and the M.S. and Ph.D. degrees in electrical engineering from the Massachusetts Institute of Technology (MIT), Cambridge, MA, USA, in 2005 and 2009, respectively. Since July 2009, he has been with Princeton University, Princeton, NJ, USA, where he is currently the Director of the Keller Center for Education in Innovation and Entrepreneurship and a Professor of electrical engineering. His research focuses on advanced sensing systems, exploring how systems for learning, inference, and action planning can be enhanced by algorithms that exploit new sensing and computing technologies. This includes research on large-area, flexible sensors, energy-efficient statistical-computing architectures and circuits, and machine-learning and statistical-signal-processing algorithms.\\
Prof. Verma was a recipient or a co-recipient of the 2006 DAC/ISSCC Student Design Contest Award, the 2008 ISSCC Jack Kilby Paper Award, the 2012 Alfred Rheinstein Junior Faculty Award, the 2013 NSF CAREER Award, the 2013 Intel Early Career Award, the 2013 Walter C. Johnson Prize for Teaching Excellence, the 2013 VLSI Symposium Best Student Paper Award, the 2014 AFOSR Young Investigator Award, the 2015 Princeton Engineering Council Excellence in Teaching Award, and the 2015 IEEE CPMT Best Paper Award. He has served as a Distinguished Lecturer for the IEEE Solid-State Circuits Society. He also serves on the technical program committees for ISSCC, VLSI Symposium, DATE, and the IEEE Signal Processing Society (DISPS).
\end{IEEEbiography}

\vspace{11pt}

\vfill

\end{document}